\documentclass[12pt,journal,draftcls,a4paper,onecolumn]{IEEEtran}

\usepackage{dsfont}
\usepackage{amsmath}
\usepackage{amssymb}
\usepackage{amsfonts}
\usepackage{graphicx}
\usepackage{amsmath}
\usepackage[noadjust]{cite}
\usepackage[all,poly]{xy}
\usepackage{multirow}

\def\bfe{{\mathbf{e}}}

\def\bfh{{\mathbf{h}}}

\def\bfx{{\mathbf{x}}}

\def\bfX{{\mathbf{X}}}
\def\bfY{{\mathbf{Y}}}

\def\dsE{{\mathds{E}}}

\def\calU{{\mathcal{U}}}

\def\calB{{\mathcal{B}}}

\def\calG{{\mathcal{G}}}

\def\calI{{\mathcal{I}}}

\def\calN{{\mathcal{N}}}

\def\calU{{\mathcal{U}}}

\newcommand{\MATobs}{\bfY}
\newcommand{\Vobs}{\mathbf{y}}
\newcommand{\obs}[1]{y_{#1}}

\newcommand{\MATima}{\bfX}
\newcommand{\Vima}{\bfx}
\newcommand{\ima}[1]{x_{#1}}

\newcommand{\dimm}{M}
\newcommand{\dimn}{P}
\newcommand{\dimima}{n}

\newcommand{\ftrans}[2]{T\left(#1,#2\right)}
\newcommand{\MATtrans}{\mathbf{H}}
\newcommand{\Vtrans}[1]{\mathbf{h}_{i}}

\newcommand{\psf}{\boldsymbol{\kappa}}

\newcommand{\MATnoise}{\mathbf{N}}
\newcommand{\Vnoise}{\mathbf{n}}
\newcommand{\noisevar}{{\sigma^2}}

\newcommand{\hypervect}{\boldsymbol{\Phi}}
\newcommand{\paramvect}{\boldsymbol{\theta}}

\newcommand{\Valpha}{\boldsymbol{\alpha}}

\newcommand{\sample}[2]{#1^{(#2)}}
\newcommand{\samplebis}[2]{{#1}^{(#2)}}
\newcommand{\samplenoisevar}[1]{{\widetilde{\sigma}}^{2(#1)}}

\newcommand{\norm}[1]{\left\|#1\right\|}

\newcommand{\R}{\mathds{R}}

\newcommand{\dirac}[1]{\delta\left({#1}\right)}

\newcommand{\transp}{^T}

\newcommand{\Vzero}{\boldsymbol{0}}
\newcommand{\Id}[1]{\textbf{I}_{#1}}
\newcommand{\Indicfun}[2]{\textbf{1}_{#1}\left(#2\right)}

\newenvironment{algogo}[1]{
\smallskip
\noindent \hrule\vspace{0.2\baselineskip} \hrule
\smallskip
\begin{small}
\refstepcounter{algo} \center{\bf \textsc{Algorithm \thealgo:}}
\\{\center{\bf #1}}
\smallskip
\flushleft
 } {
\end{small}
\bigskip
\hrule\vspace{0.2\baselineskip} \hrule
\smallskip }

\newcounter{algo}
\renewcommand{\thealgo}{\arabic{algo}}

\newcommand{\figwidth}{\columnwidth}

\title{Hierarchical Bayesian sparse image \\reconstruction with application to MRFM}

\author{Nicolas Dobigeon$^{1,2}$, Alfred O. Hero$^{2}$ and Jean-Yves Tourneret$^{1}$
\\
\normalsize $^1$ University of Toulouse, IRIT/INP-ENSEEIHT, 2 rue Camichel, 31071 Toulouse, France. \\
\normalsize $^2$ University of Michigan, Department of EECS, Ann Arbor, MI 48109-2122, USA \\
\small\texttt{\{Nicolas.Dobigeon, Jean-Yves.Tourneret\}@enseeiht.fr,
hero@umich.edu}
\thanks{Part of this work has been supported by ARO MURI grant No. W911NF-05-1-0403.}}

\begin{document}

\maketitle

\hyphenation{hie-rar-chi-cal}

\begin{abstract}
This paper presents a hierarchical Bayesian model to reconstruct
sparse images when the observations are obtained from linear
transformations and corrupted by an additive white Gaussian noise.
Our hierarchical Bayes model is well suited to such naturally sparse
image applications as it seamlessly accounts for properties such as
sparsity and positivity of the image via appropriate Bayes priors.
We propose a prior that is based on a weighted mixture of a positive
exponential distribution and a mass at zero. The prior has
hyperparameters that are tuned automatically by marginalization over
the hierarchical Bayesian model. To overcome the complexity of the
posterior distribution, a Gibbs sampling strategy is proposed. The
Gibbs samples can be used to estimate the image to be recovered,
e.g. by maximizing the estimated  posterior distribution. In our
fully Bayesian approach the posteriors of all the parameters are
available. Thus our algorithm provides more information than other
previously proposed sparse reconstruction methods that only give a
point estimate. The performance of the proposed hierarchical
Bayesian sparse reconstruction method is illustrated on synthetic
data and real data collected from a tobacco virus sample using a
prototype MRFM instrument.
\end{abstract}

\begin{keywords}
Deconvolution, MRFM imaging, sparse representation, Bayesian
inference, MCMC methods.
\end{keywords}

\newpage

\section{Introduction}
\label{sec:intro} For several decades, image deconvolution has been
of increasing interest \cite{Andrews1977,Russ2006}. Image
deconvolution is a method for reconstructing images from
observations provided by optical or other devices and may include
denoising, deblurring or restoration. The applications are numerous
including astronomy \cite{Starck2006}, medical imagery
\cite{Sarder2006}, remote sensing \cite{Reichenbach1995} and
photography \cite{Sroubek2003}. More recently, a new imaging
technology, called Magnetic Resonance Force Microscopy (MRFM), has
been developed (see \cite{Mounce2004} and \cite{Kuehn2008} for
reviews). This non-destructive method allows one to improve the
detection sensitivity of standard magnetic resonance imaging (MRI)
\cite{Rugar2004}. Three dimensional MRI at $4$nm spatial resolution
has recently been achieved by the IBM MRFM prototype for imaging the
proton density of a tobacco virus \cite{Degen2009}. Because of its
potential atomic-level resolution\footnote{Note that the current
state of art of the MRFM technology allows one to acquire images
with nanoscale resolution. However, atomic-level resolution might be
obtained in the future.}, the $2$-dimensional or $3$-dimensional
images resulting from this technology are naturally sparse in the
standard pixel basis. Indeed, as the observed objects are molecules,
most of the image is empty space. In this paper, a hierarchical
Bayesian model is proposed to perform reconstruction of such images.

Sparse signal and image deconvolution has motivated research in many
scientific applications including: spectral analysis in astronomy
\cite{Bourguignon2007}; seismic signal analysis in geophysics
\cite{Cheng1996,Rosec2003}; and deconvolution of ultrasonic B-scans
\cite{Olofsson2007}. We propose here a hierarchical Bayesian model
that is based on selecting an appropriate prior distribution for the
unknown image and other unknown parameters. The image prior is
composed of a weighted mixture of a standard exponential
distribution and a mass at zero. When the non-zero part of this
prior is chosen to be a centered normal distribution, this prior
reduces to a Bernoulli-Gaussian process. This distribution has been
widely used in the literature to build Bayesian estimators for
sparse deconvolution problems (see
\cite{Kormylo1982,Idier1990,Lavielle1993,Champagnat1996,Doucet1997}
or more recently \cite{Bourguignon2005ssp} and \cite{Fevotte2008}).
However, choosing a distribution with heavier tail may improve the
sparsity inducement of the prior. Combining a Laplacian distribution
with an atom at zero results in the so-called LAZE prior. This
distribution has been used in \cite{Johnstone2004} to solve a
general denoising problem in a non-Bayesian quasi-maximum likelihood
estimation framework. In \cite{Ting2006icip,Ting2006}, this prior
has also been used for sparse reconstruction of noisy images,
including MRFM. The principal weakness of these previous approaches
is the sensitivity to hyperparameters that determine the prior
distribution, e.g. the LAZE mixture coefficient and the weighting of
the prior vs the likelihood function. The hierarchical Bayesian
approach proposed in this paper circumvents these difficulties.
Specifically, a new prior composed of a mass at zero and a
single-sided exponential distribution is introduced, which accounts
for positivity and sparsity of the pixels in the image. Conjugate
priors on the hyperparameters of the image prior are introduced. It
is this step that makes our approach hierarchical Bayesian. The full
Bayesian posterior can then be derived from samples generated by
Markov chain Monte Carlo (MCMC) methods \cite{Robert2004}.

The estimation of hyperparameters involved in the prior distribution
described above is the most difficult task  and poor estimation
leads to instability. Empirical Bayes (EB) and Stein unbiased risk
(SURE) solutions were proposed in \cite{Ting2006,Ting2006icip} to
deal with this issue. However, instability was observed especially
at higher signal-to-noise ratios (SNR). In the Bayesian estimation
framework, two approaches are available to estimate these
hyperparameters. One approach couples MCMC methods to an
expectation-maximization (EM) algorithm or to a stochastic EM
algorithm \cite{Lavielle2001,Kuhn2004} to maximize a penalized
likelihood function. The second approach defines non-informative
prior distributions for the hyperparameters; introducing a second
level of hierarchy into the Bayesian formulation. This latter fully
Bayesian approach, adopted in this paper, has been successfully
applied to signal segmentation
\cite{Dobigeon2007a,Dobigeon2007b,Dobigeon2007c} and semi-supervised
unmixing of hyperspectral imagery \cite{Dobigeon2008}.

Only a few papers have been published on reconstruction of images
from MRFM data \cite{Chao2004,Zuger1994,Zuger1996,Degen2009}. In
\cite{Hammel2003}, several techniques based on linear filtering and
maximum-likelihood principles were proposed that do not exploit
image sparsity. More recently, Ting \emph{et al.} has introduced
sparsity penalized reconstruction methods for MRFM (see
\cite{Ting2006} or \cite{Ting2009}). The reconstruction problem has
been formulated as a decomposition into a deconvolution step and a
denoising step, yielding an iterative thresholding framework. In
\cite{Ting2006} the hyperparameters are estimated using penalized
log-likelihood criteria including the SURE approach
\cite{Stein1981}. Despite promising results, especially at low SNR,
penalized likelihood approaches require iterative maximization
algorithms that are often slow to converge and can get stuck on
local maxima \cite{Diebolt1996}. In contrast to \cite{Ting2006}, the
fully Bayesian approach presented in this paper converges quickly
and produces estimates of the entire posterior and not just the
local maxima. Indeed, the hierarchical Bayesian formulation proposed
here generates Bayes-optimal estimates of all image parameters,
including the hyperparameters.

In this paper, the response of the MRFM imaging device is assumed to
be known. While it may be possible to extend our methods to unknown
point spread functions, e.g., along the lines of
\cite{Herrity2008,Herrity2008b}, the case of sparse blind
deconvolution is outside of the scope of this paper.

This paper is organized as follows. The deconvolution problem is
formulated in Section~\ref{sec:problem}. The hierarchical Bayesian
model is described in Section~\ref{sec:model}.
Section~\ref{sec:Gibbs} presents a Gibbs sampler that allows one to
generate samples distributed according to the posterior of interest.
Simulation results, including extensive performance comparison, are
presented in Section~\ref{sec:simu}. In Section \ref{sec:simu_real}
we apply our hierarchical Bayesian method to reconstruction of a
tobacco virus from real MRFM data. Our main conclusions are reported
in Section~\ref{sec:conclusions}.

\section{Problem formulation} \label{sec:problem}
Let $\MATima$ denote a $l_1\times \ldots \times l_{\dimima}$ unknown
$\dimima$-dimensional pixelated image to be recovered (e.g.
$\dimima=2$ or $\dimima=3$). Observed are a collection of $\dimn$
projections $\Vobs=\left[\obs{1},\ldots,\obs{\dimn}\right]\transp$
which are assumed to follow the model:
\begin{equation}
\label{eq:model_nD}
  \Vobs = \ftrans{\boldsymbol{\kappa}}{\MATima} + \Vnoise,
\end{equation}
where $\ftrans{\cdot}{\cdot}$ stands for a bilinear function,
$\Vnoise$ is a $\dimn\times 1$ dimension noise vector and
$\boldsymbol{\kappa}$ is the kernel that characterizes the response
of the imaging device. In the right-hand side of
\eqref{eq:model_nD}, $\Vnoise$ is an additive Gaussian noise
sequence distributed according to $\Vnoise \sim
\calN\left(\Vzero,\noisevar\Id{\dimn}\right)$, where the variance
$\noisevar$ is assumed to be unknown.

Note that in standard deblurring problems, the function
$\ftrans{\cdot}{\cdot}$ represents the standard
$\dimima$-dimensional convolution operator $\otimes$. In this case,
the image $\MATima$ can be vectorized yielding the unknown image
$\Vima \in \R^\dimm$ with $\dimm = \dimn = l_1 l_2\ldots l_\dimima$.
With this notation, Eq.~\eqref{eq:model_nD} can be rewritten:
\begin{equation}    \label{eq:model}
  \Vobs        = \MATtrans\Vima + \Vnoise \qquad \text{or} \qquad \MATobs = \boldsymbol{\kappa} \otimes \MATima + \MATnoise
\end{equation}
where $\Vobs$ (resp. $\Vnoise$) stands for the vectorized version of
$\MATobs$ (resp. $\MATnoise$) and $\MATtrans$ is an $\dimn \times
\dimm$ matrix that describes convolution by the point spread
function (psf) $\boldsymbol{\kappa}$.

The problem addressed in the following sections consists of
estimating $\Vima$ and $\noisevar$ under sparsity and positivity
constraints on $\Vima$ given the observations $\Vobs$, the psf
$\boldsymbol{\kappa}$ and the bilinear function\footnote{In the
following, for sake of conciseness, the same notation
$\ftrans{\cdot}{\cdot}$ will be adopted for the bilinear operations
used on $\dimima$-dimensional images $\MATima$ and used on $\dimm
\times 1$ vectorized images $\Vima$.} $\ftrans{\cdot}{\cdot}$.

\section{Hierarchical Bayesian model}
\label{sec:model}
\subsection{Likelihood function}
The observation model defined in \eqref{eq:model_nD} and the
Gaussian properties of the noise sequence $\Vnoise$ yield:
\begin{equation}
  \label{eq:likelihood}
  f\left(\Vobs | \Vima, \noisevar\right) =
  \left(\frac{1}{2\pi\sigma^2}\right)^{\frac{\dimn}{2}}
  \exp\left(-\frac{\norm{\Vobs-\ftrans{\boldsymbol{\kappa}}{\Vima}}^2}{2\noisevar}\right),
\end{equation}
where $\norm{\cdot}$ denotes the standard $\ell_2$ norm:
$\norm{\bfx}^2=\bfx\transp\bfx$.

\subsection{Parameter prior distributions}

The unknown parameter vector associated with the observation model
defined in \eqref{eq:model_nD} is $\paramvect=\left\{\Vima,
\noisevar\right\}$. In this section, we introduce prior
distributions for these two parameters; which are assumed to be
independent.

\subsubsection{Image prior}
\label{subsubsec:prior_ima} First consider the exponential
distribution with shape parameter $a>0$:
\begin{equation}
  \label{eq:prior_ima}
  g_a\left(\ima{i}\right) =
  \frac{1}{a}\exp\left(-\frac{\ima{i}}{a}\right)\Indicfun{\R_+^*}{\ima{i}},
\end{equation}
where $\Indicfun{\dsE}{x}$ is the indicator function defined on
$\dsE$:
\begin{equation}
  \Indicfun{\dsE}{x}=\left\{
                      \begin{array}{ll}
                        1, & \hbox{if $x\in\dsE$,} \\
                        0, & \hbox{otherwise.}
                      \end{array}
                    \right.
\end{equation}
Choosing $g_a\left(\cdot\right)$ as prior distributions for
$\ima{i}$ ($i=1,\ldots,\dimm$) leads to a maximum \emph{a
posteriori} (MAP) estimator of $\Vima$ that corresponds to a maximum
$\ell_1$-penalized likelihood estimate with a positivity
constraint\footnote{Note that a similar estimator using a Laplacian
prior for $x_i$ ($i=1,\ldots,\dimm$) was proposed in
\cite{Tibshirani1996} for regression problems, referred to as the
LASSO estimator, but without positivity constraint.}. Indeed,
assuming the component $\ima{i}$ ($i=1,\ldots,\dimn$) a priori
independent allows one to write the full prior distribution for
$\Vima=\left[\ima{1},\ldots,\ima{\dimm}\right]\transp$:
\begin{equation}
  g_a\left(\Vima\right) = \left(\frac{1}{a}\right)^{\dimm}
\exp\left(-\frac{\left\|\Vima\right\|_1}{a}\right)
\Indicfun{\left\{\Vima\succ0\right\}}{\Vima},
\end{equation}
where $\left\{\Vima\succ0\right\} = \left\{\Vima \in\R^\dimm ;
\ima{i}>0, \forall i=1,\ldots,\dimm\right\}$ and
$\left\|\cdot\right\|_1$ is the standard $\ell_1$ norm
$\left\|\bfx\right\|_1=\sum_i |x_i|$. This estimator has interesting
sparseness properties for Bayesian estimation \cite{Alliney1994} and
signal representation \cite{Gribonval2003}.

Coupling a standard probability density function (pdf) with an atom
at zero is another alternative to encourage sparsity. This strategy
has for instance been used for located event detection
\cite{Kormylo1982} such as spike train deconvolution
\cite{Cheng1996,Champagnat1996}. In order to increase the sparsity
of the prior, we propose to use the following distribution derived
from $g_a\left(\cdot\right)$ as prior distribution for $\ima{i}$:
\begin{equation}
\label{eq:prior_ima2}
  f\left(\ima{i}|w,a\right) = (1-w) \dirac{\ima{i}} + w
g_a\left(\ima{i}\right),
\end{equation}
where $\delta\left(\cdot\right)$ is the Dirac function. This prior
is similar to the LAZE distribution (Laplacian pdf and an atom at
zero) introduced in \cite{Johnstone2004} and used, for example, in
\cite{Ting2006,Ting2006icip} for MRFM. However, since $g_a(\ima{i})$
is zero for $\ima{i}\leq0$, the proposed prior in
\eqref{eq:prior_ima2} accounts for the positivity of the non-zero
pixel values, a constraint that exists in many imaging modalities
such as MRFM. By assuming the components $\ima{i}$ to be a priori
independent ($i=1,\ldots,\dimm$), the following prior distribution
for $\Vima$ is obtained:
\begin{equation}
  f\left(\Vima|w,a\right) = \prod_{i=1}^\dimm\left[(1-w) \delta\left(\ima{i}\right) + w
g_a\left(\ima{i}\right)\right].
\end{equation}

Introducing the index subsets $\calI_0=\left\{i; \ima{i}=0\right\}$
and  $\calI_1=\overline{\calI}_0=\left\{i; \ima{i}\neq0\right\}$
allows one to rewrite the previous equation as follows:
\begin{equation}
    \label{eq:prior_Vima}
  f\left(\Vima|w,a\right) = \left[\left(1-w\right)^{n_0}\prod_{i\in\calI_0} \delta\left(\ima{i}\right)\right]
            \left[ w^{n_1}\prod_{i\in\calI_1}g_a\left(\ima{i}\right)\right],
\end{equation}
with $n_\epsilon = \mathrm{card}\left\{\calI_\epsilon\right\}$,
$\epsilon\in\left\{0,1\right\}$. Note that $n_0=\dimm - n_1$ and
$n_1= \left\|\Vima\right\|_0$ where $\left\|\cdot\right\|_0$ is the
standard $\ell_0$ norm $\left\|\Vima\right\|_0 = \#\left\{i;
\ima{i}\neq0\right\}$.

\subsubsection{Noise variance prior} A conjugate
inverse-Gamma distribution with parameters $\frac{\nu}{2}$ and
$\frac{\gamma}{2}$ is chosen as prior distribution for the noise
variance \cite[Appendix A]{Robert2001}:
\begin{equation}
    \label{eq:prior_noisevar}
  \noisevar | \nu,\gamma\sim
\calI\calG\left(\frac{\nu}{2},\frac{\gamma}{2}\right).
\end{equation}
In the following, the shape parameter $\nu$ will be fixed to $\nu=2$
and the scale parameter $\gamma$ will be estimated as an
hyperparameter (see \cite{Punskaya2002,Dobigeon2007b,Dobigeon2008}).
Note that choosing the inverse-Gamma distribution $\calI
\calG\left(\frac{\nu}{2},\frac{\gamma}{2}\right)$ as a prior for
$\noisevar$ is equivalent to choosing a Gamma distribution
$\calG\left(\frac{\nu}{2},\frac{\gamma}{2}\right)$ as a prior for
$1/\sigma^2$.

\subsection{Hyperparameter priors}
The hyperparameter vector associated with the aforementioned prior
distributions is $\hypervect=\left\{a,\gamma,w\right\}$. Obviously,
the accuracy of the proposed Bayesian model depends on the values of
these hyperparameters. Sometimes prior knowledge may be available,
e.g., the mean number of non-zero pixels  in the image. In this case
these parameters can be tuned manually to their true values.
However, in many practical situations such prior information is not
available and these hyperparameters must be estimated directly from
the data. Priors for these hyperparameters, sometimes referred to as
``hyperpriors" are given below.

\subsubsection{Hyperparameter $a$} A conjugate inverse-Gamma
distribution is assumed for the scale parameter $a$ of the
distribution $g_a\left(\cdot\right)$ of non-zero pixel intensities:
\begin{equation}
  a | \Valpha \sim \calI\calG\left(\alpha_0,\alpha_1\right),
\end{equation}
with $\Valpha=\left[\alpha_0,\alpha_1\right]\transp$. Similarly to
\cite{Godsill1998}, the fixed hyperparameters $\alpha_0$ and
$\alpha_1$ have been chosen to obtain a vague prior: $\alpha_0 =
\alpha_1 = 10^{-10}$.

\subsubsection{Hyperparameter $\gamma$} A non informative
Jeffreys' prior \cite{Jeffreys1946,Jeffreys1961} is assumed for the
scale parameter of the inverse Gamma prior density on the noise
variance $\noisevar$  :
\begin{equation}
   \label{eq:prior_gamma}
    f\left(\gamma\right)\propto\frac{1}{\gamma} \Indicfun{\R_+}{\gamma}.
\end{equation}
The combination of the priors \eqref{eq:prior_noisevar} and
\eqref{eq:prior_gamma} yields the non-informative Jeffreys' prior on
$\noisevar$.  Note that there is no difference between choosing a
non-informative Jeffrey's prior for $\noisevar$ and the proper
hierarchical prior defined by \eqref{eq:prior_noisevar} and
\eqref{eq:prior_gamma}. Indeed, integrating over the hyperparameter
$\gamma$ in the joint $f\left(\noisevar,\gamma\right)$ distribution
yields:
\begin{equation}
\begin{split}
  f\left(\noisevar\right)
    &= \int f\left(\noisevar|\gamma\right)f\left(\gamma\right)d\gamma\\
    &\propto \left(\frac{1}{\noisevar}\right)^{2} \int
    \exp\left(-\frac{\gamma}{2\noisevar}\right)d\gamma\\
    &\propto \frac{1}{\noisevar}.
\end{split}
\end{equation}
However, in more complex noise models the hierarchical priors
$f\left(\noisevar|\gamma\right)$ and $f\left(\gamma\right)$ are not
equivalent to such a simple prior on $\noisevar$. For example,  as
in \cite{DobigeonTR2008b}, this pair of hierarchical priors is
easily generalizable to  conditionally Gaussian noise with spatial
correlation and spatially varying signal-to-noise ratio.

\subsubsection{Hyperparameter $w$}
A uniform distribution on the simplex $[0,1]$ has been chosen as
prior distribution for the mean proportion of non-zero pixels:
\begin{equation}
  w \sim \calU\left([0,1]\right).
\end{equation}
This is the least informative prior on the image sparsity factor.
Assuming that the individual hyperparameters are statistically
independent the full hyperparameter prior distribution for
$\hypervect$  can be expressed as:
\begin{equation}
\begin{split}
\label{eq:prior_Hyper}
  f\left(\hypervect|\Valpha\right) &=
f\left(w\right)f\left(\gamma\right)f\left(a\right) \\
&\propto\frac{1}{
 \gamma a^{\alpha_0+1}
}\exp\left(-\frac{\alpha_1}{a}\right)\\
&\times\Indicfun{[0,1]}{w} \Indicfun{\mathbb{R}^+}{a}
\Indicfun{\mathbb{R}^+}{\gamma},
\end{split}
\end{equation}

\subsection{Posterior distribution}

\begin{figure}[t!]
  \centering
  \includegraphics[width=\figwidth]{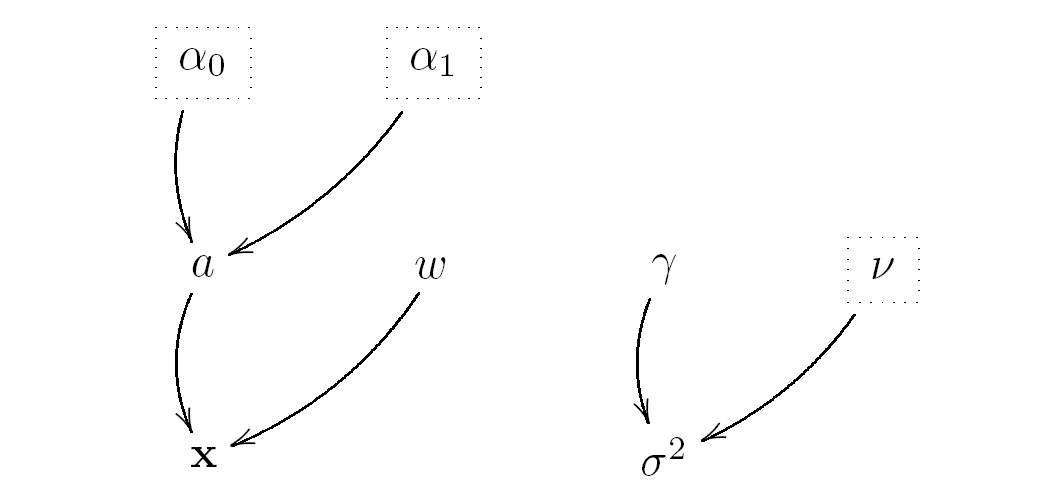}
  \caption{DAG for the parameter priors and hyperpriors (the fixed non-random hyperparameters appear in dashed boxes).}\label{fig:DAG}
\end{figure}

The posterior distribution of $\left\{\paramvect,\hypervect\right\}$
can be computed as follows:
\begin{equation}
\label{eq:fullposterior}
f\left(\paramvect,\hypervect|\Vobs,\Valpha\right) \propto
f\left(\Vobs|\paramvect\right)f\left(\paramvect|\hypervect\right)f\left(\hypervect|\Valpha\right),
\end{equation}
with
\begin{equation}
f\left(\paramvect|\hypervect\right) = f\left(\Vima|
a,w\right)f\left(\sigma^2| \gamma\right),
\end{equation}
where $f\left(\Vobs|\paramvect\right)$ and
$f\left(\hypervect|\Valpha\right)$ have been defined in
\eqref{eq:likelihood} and \eqref{eq:prior_Hyper}. This hierarchical
structure, represented on the directed acyclic graph (DAG) in
Fig.~\ref{fig:DAG}, allows one to integrate out the parameter
$\noisevar$ and the hyperparameter vector $\hypervect$
in~\eqref{eq:fullposterior} to obtain the posterior of the image
given the measured data and the parameters $\Vima$:
\begin{equation}
\begin{split}
 \label{eq:posterior}
  f\left(\Vima|\Vobs,\Valpha\right) \propto
 \frac{B\left(1+ n_1, 1 +
n_0\right)}{\left\|\Vobs-\ftrans{\boldsymbol{\kappa}}{\Vima}\right\|^{\dimn}}\\
    \times \frac{\Gamma\left(n_1+\alpha_0\right)}{\left[\norm{\Vima}_1+\alpha_1\right]^{n_1+\alpha_0}} \Indicfun{\left\{\Vima\succ0\right\}}{\Vima}.
\end{split}
\end{equation}
In \eqref{eq:posterior}, as defined in
paragraph~\ref{subsubsec:prior_ima}, $n_1 = \norm{\Vima}_0$, $n_0 =
\dimm -\norm{\Vima}_0$ and $B\left(\cdot,\cdot\right)$ stands for
the Beta function $B\left(u,v\right)=
\Gamma\left(u\right)\Gamma\left(v\right) \slash
{\Gamma\left(u+v\right)}$, where $\Gamma(\cdot)$ denotes the Gamma
function.

The next section presents an appropriate Gibbs sampling strategy
\cite[Chap. 10]{Robert2004} that allows one to generate an image
sample distributed according to the posterior distribution
$f\left(\Vima|\Vobs,\Valpha\right)$.

\section{A Gibbs sampling strategy \\for sparse image reconstruction}
\label{sec:Gibbs} In this section, we describe the Gibbs sampling
strategy for generating samples
$\left\{\sample{\Vima}{t}\right\}_{t=1,\ldots}$ distributed
according to the posterior distribution in \eqref{eq:posterior}. As
simulating directly according to \eqref{eq:posterior} is difficult,
it is much more convenient to generate samples distributed according
to the joint posterior
$f\left(\Vima,\noisevar,a,w|\Vobs,\Valpha\right)$. This Gibbs
sampler produces sequences
$\left\{{\Vima}^{(t)}\right\}_{t=1,\ldots}$,
$\left\{{\sigma}^{2(t)}\right\}_{t=1,\ldots}$,
$\left\{{a}^{(t)}\right\}_{t=1,\ldots}$,
$\left\{{w}^{(t)}\right\}_{t=1,\ldots}$ which are Markov chains with
stationary distributions $f\left(\Vima|\Vobs,\Valpha\right)$,
$f\left(\noisevar|\Vobs,\Valpha\right)$,
$f\left(a|\Vobs,\Valpha\right)$ and $f\left(w|\Vobs,\Valpha\right)$,
respectively \cite[p. 345]{Robert2004}. Then, the MAP estimator of
the unknown image $\Vima$ will be computed by retaining among
$\mathcal{X} = \left\{\sample{\Vima}{t} \right\}_{t=1,\ldots}$ the
generated sample that maximizes the posterior distribution in
\eqref{eq:posterior} \cite[p. 165]{Marin2007}:
\begin{equation} \label{eq:MAP_estimate}
\begin{split} \hat\Vima_{\text{MAP}} &=
    \operatornamewithlimits{argmax}_{\Vima\in \mathbb{R}_+^{M}}
    f\left(\Vima|\Vobs\right)\\
    & \approx \operatornamewithlimits{argmax}_{\Vima\in
\mathcal{X}}
    f\left(\Vima|\Vobs\right).
\end{split}
\end{equation}
The main steps of this algorithm are given in
subsections~\ref{subsec:sample_w} and~\ref{subsec:sample_noisevar}
(see also Algorithm~\ref{algo:Gibbs} below).

\begin{figure}[h!]
\begin{algogo}{Gibbs sampling algorithm  for sparse image reconstruction}
    \label{algo:Gibbs}
    \begin{itemize}
    \item \underline{Initialization:}
    \begin{itemize}
        \item Sample parameter $\sample{\Vima}{0}$ from the pdf in
            \eqref{eq:prior_Vima},
        \item Sample parameter $\samplenoisevar{0}$ from the pdf in \eqref{eq:prior_noisevar},
        \item Set $t \leftarrow 1$,
    \end{itemize}
    \item \underline{Iterations:} for $t=1,2, \ldots, $ do
    \begin{itemize}
        \item[1.] Sample hyperparameter $\sample{w}{t}$ from the pdf in
            \eqref{eq:posterior_w},
        \item[2.] Sample hyperparameter $\sample{a}{t}$ from the pdf in
            \eqref{eq:posterior_a},
        \item[3.]  \label{algostep:sample_ima}For $i=1,\ldots,\dimm$, sample parameter $\sample{\ima{i}}{t}$ from the pdf in
            \eqref{eq:posterior_ima},
        \item[4.] Sample parameter $\samplenoisevar{t}$ from the pdf in \eqref{eq:posterior_noisevar},
        \item[5.] Set $t \leftarrow t+1$.
    \end{itemize}
    \end{itemize}
\end{algogo}
\end{figure}

\subsection{Generation of samples according to
$f\left(w\left|\Vima\right.\right)$} \label{subsec:sample_w}
 Using \eqref{eq:prior_Vima}, the
following result can be obtained:
\begin{equation}
  f\left(w\left|\Vima\right.\right) \propto
(1-w)^{n_0}w^{n_1},
\end{equation}
where $n_0$ and $n_1$ have been defined in paragraph
\ref{subsubsec:prior_ima}. Therefore, samples from
$f\left(w\left|\Vima\right.\right)$  can be generated by simulating
from an image dependent Beta distribution:
\begin{equation}
\label{eq:posterior_w}
  w\left|\Vima\right. \sim \calB e\left(1 + n_1, 1 +
n_0\right).
\end{equation}

\subsection{Generation of samples according to $f\left(a\left|\Vima,\Valpha\right.\right)$}
The form of  the joint posterior distribution
\eqref{eq:fullposterior} implies that samples of $a$ can be
generated by simulating from an image dependent inverse-Gamma
distribution:
\begin{equation}
\label{eq:posterior_a}
  a\left|\Vima,\Valpha\right. \sim
\calI\calG\left(\norm{\Vima}_0 + \alpha_0,\norm{\Vima}_1+
\alpha_1\right).
\end{equation}

\subsection{Generation of samples according to
$f\left(\Vima\left|w,a,\noisevar,\Vobs\right.\right)$} The LAZE-type
prior \eqref{eq:prior_ima2} chosen for $\ima{i}$
($i=1,\ldots,\dimm$) yields a posterior distribution of $\Vima$ that
is not closed form. However, one can easily derive the posterior
distribution of each pixel intensity $\ima{i}$ ($i=1,\ldots,\dimm$)
conditioned on the intensities of the rest of the image. Indeed
straightforward computations (Appendix \ref{app:derivation1}) yield:
\begin{equation}
\begin{split}
    \label{eq:posterior_ima}
  f\left(\ima{i} | w,a,\noisevar,\Vima_{-i},\Vobs\right) &\propto
(1-w_i) \delta\left(\ima{i}\right) \\
&+ w_i \phi_+\left(\ima{i}|\mu_i,\eta^2_i\right),
\end{split}
\end{equation}
where $\Vima_{-i}$ stands for the vector $\Vima$ whose $i$th
component has been removed and $\mu_i$ and $\eta^2_i$ are given in
Appendix \ref{app:derivation1}. In \eqref{eq:posterior_ima},
$\phi_+\left(\cdot,m,s^2\right)$ stands for the pdf of the truncated
Gaussian distribution defined on $\R^*_+$ with hidden mean and
variance parameters equal to $m$ and $s^2$, respectively:
\begin{equation}
  \phi_+\left(x,m,s^2\right) = \frac{1}{C\left(m,s^2\right)}
\exp\left[-\frac{\left(x-m\right)^2}{2s^2}\right]\Indicfun{\R^*_+}{x},
\end{equation}
with
\begin{equation}
    \label{eq:constant_C}
  C\left(m,s^2\right) = \sqrt{\frac{\pi
            s^2}{2}} \left[1 +
\mathrm{erf}\left(\frac{m}{\sqrt{2s^2}}\right)\right].
\end{equation}
The form in \eqref{eq:posterior_ima} specifies $\ima{i} |
w,a,\noisevar,\Vima_{-i},\Vobs$ as a Bernoulli-truncated Gaussian
variable with parameter $\left(w_i, \mu_i, \eta^2_i\right)$.
Appendix~\ref{app:Gene_BeTrG} presents an algorithm that can be used
to generate samples from this distribution. This algorithm generates
samples distributed according to $f\left(\Vima\left|w,
\noisevar,a,\Vobs\right.\right)$ by successively updating the
coordinates of $\Vima$ using a sequence of $\dimm$ Gibbs moves
(requiring generation of Bernoulli-truncated Gaussian variables).

\subsection{Generation of samples according to $f\left(\noisevar\left|\Vima,\Vobs\right.\right)$}
\label{subsec:sample_noisevar} Samples are generated in the
following way:
\begin{equation}
    \label{eq:posterior_noisevar}
  \noisevar\left|\Vima,\Vobs\right. \sim
\calI\calG\left(\frac{\dimn}{2},\frac{\norm{\Vobs -
\ftrans{\boldsymbol{\kappa}}{\Vima}}^2}{2}\right).
\end{equation}

\section{Simulation on synthetic images}
\label{sec:simu}

\begin{table}[h!]
\renewcommand{\arraystretch}{1.4}
\begin{center}
\caption{Parameters used to compute the MRFM
psf.\label{tab:parameters}}
\begin{tabular}{|l|c|c|}
\hline
\multicolumn{2}{|c|}{Parameter} & \multirow{2}{5mm}{Value}\\
\cline{1-2}
Description & Name  & \\
\hline
Amplitude of external magnetic field            & $B_\textrm{ext}$    & $9.4 \times 10^3~\textrm{G}$ \\
Value of $B_\textrm{mag}$ in the resonant slice & $B_\textrm{res}$    & $1.0 \times 10^4~\textrm{G}$ \\
Radius of tip                                   & $R_0$               & $4.0~\textrm{nm}$ \\
Distance from tip to sample                     & $d$                 & $6.0~\textrm{nm}$ \\
Cantilever tip moment                           & $m$                 & $4.6 \times 10^5~\textrm{emu}$\\
Peak cantilever oscillation         & $x_\textrm{pk}$     & $0.8~\textrm{nm}$\\
Maximum magnetic field gradient                 & $G_\textrm{max}$    & $125$ \\
\hline
\end{tabular}
\end{center}
\end{table}

\begin{figure}[h!]
  \centering
  \includegraphics[width=\figwidth]{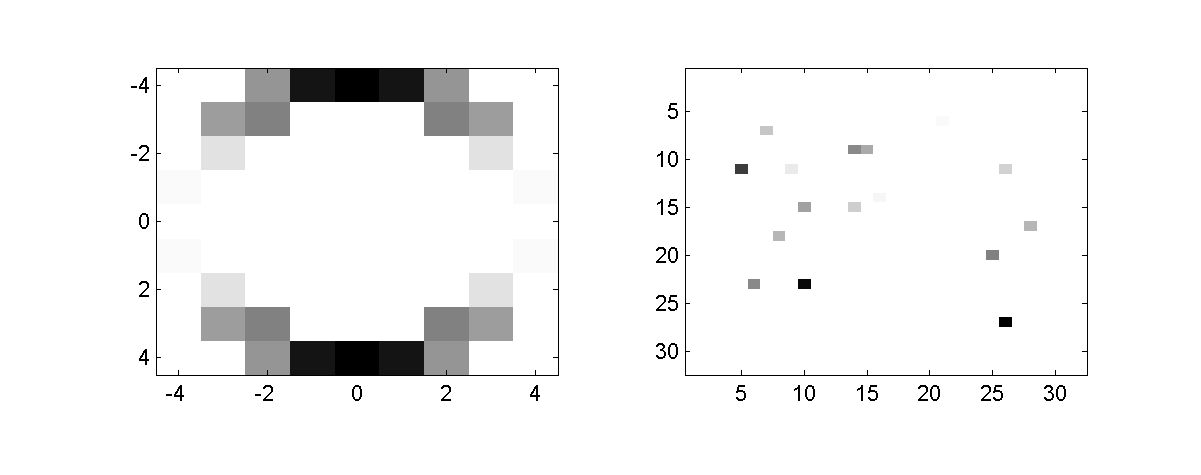}
  \caption{Left: Psf of the MRFM tip. Right: unknown sparse image to be estimated.}
  \label{fig:image_synth}
\end{figure}

\subsection{Reconstruction of 2-dimensional image}
In this subsection, a $32 \times 32$ synthetic image, depicted in
Fig.~\ref{fig:image_synth}~(right panel), is simulated using the
prior in \eqref{eq:prior_Vima} with parameters $a=1$ and $w=0.02$.
In Figs. \ref{fig:image_synth} and \ref{fig:image_synth_result},
white pixels stands for zero intensity values. A general analytical
derivation of the psf of the MRFM tip has been given in
\cite{Mamin2003} and with further explanation in \cite{Ting2006}.
Following this model, we defined a $10 \times 10$ $2$-dimensional
convolution kernel, the psf represented in
Fig.~\ref{fig:image_synth} (left panel), that corresponds to the
physical parameters shown in Table~\ref{tab:parameters}. The
associated psf  matrix $\MATtrans$ introduced in \eqref{eq:model} is
of size $1024 \times 1024$. The observed measurements $\Vobs$, which
are of size $\dimn=1024$ and depicted in Fig.
\ref{fig:image_synth_result} (top panel), are corrupted by an
additive Gaussian noise with two different variances $\noisevar=1.2
\times 10^{-1}$ and $\noisevar=1.6 \times 10^{-3}$, corresponding to
signal-to-noise ratios $\mathrm{SNR} = 2$dB and $\mathrm{SNR} =
20$dB, respectively.

\subsubsection{Simulation results} \label{subsec:simu_results} 
\begin{figure}[h!]
  \centering
  \includegraphics[width=\figwidth]{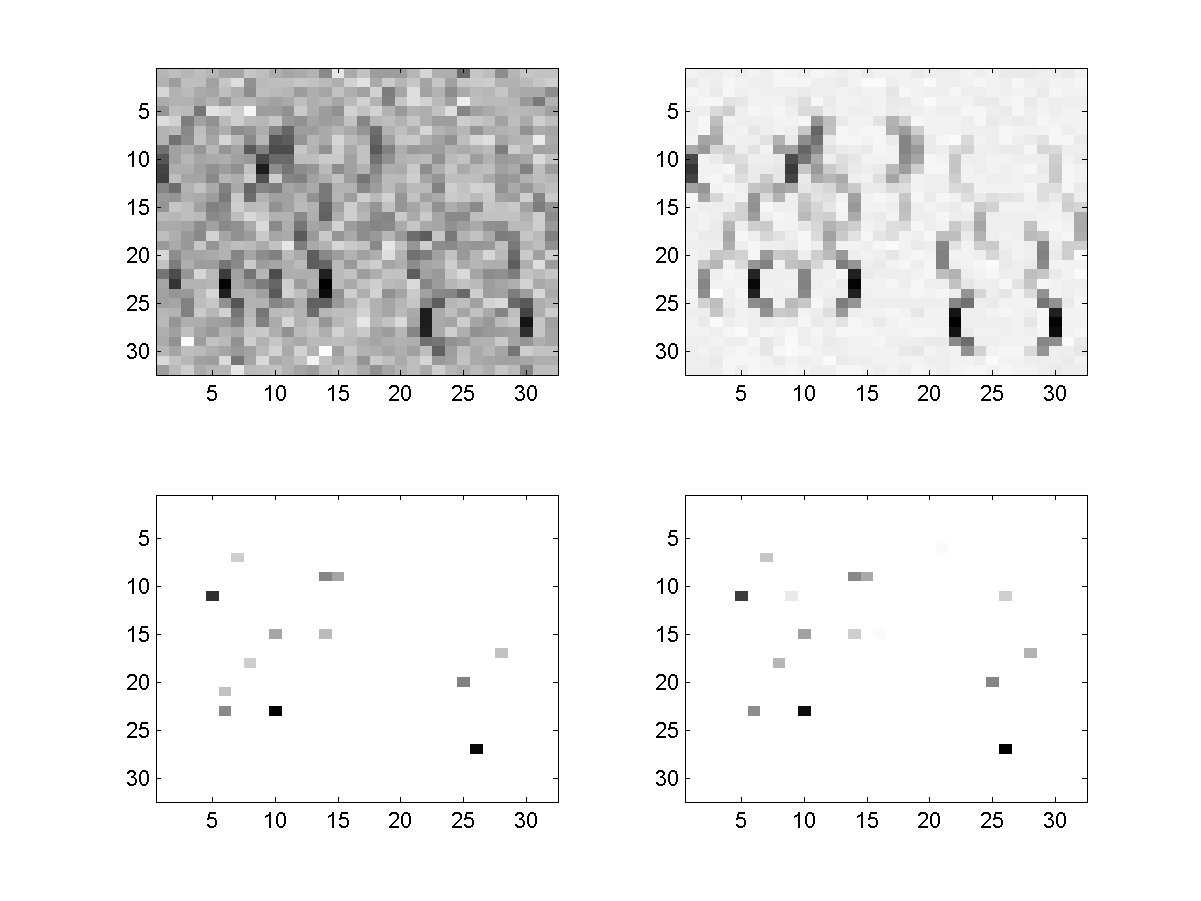}
  \caption{Top, left (resp. right): noisy observations for $\textrm{SNR}=2$dB (resp. $20$dB).
    Bottom, left (resp. right): reconstructed image for $\textrm{SNR}=2$dB (resp. $20$dB).}
  \label{fig:image_synth_result}
\end{figure}

The
observations are processed by the proposed algorithm using
$N_\textrm{MC} = 2000$ iterations of the Gibbs sampler with
$N_\textrm{bi}=300$ burn-in iterations. The computation time for
completing $100$ iterations of the proposed algorithm is $80$s for
an unoptimized MATLAB 2007b 32bit implementation on a 2.2GHz Intel
Core 2, while $100$ iterations of the Landweber and empirical
Bayesian algorithms require $0.15$s and $2$s, respectively. The MAP
image reconstruction computed using \eqref{eq:MAP_estimate} is
depicted in Fig.~\ref{fig:image_synth_result} (bottom panel) for the
two levels of noise considered. Observe that the estimated image is
very similar to the actual image, Fig. \ref{fig:image_synth} (right
panel), even at low SNR.

\begin{figure}[h!]
  \centering
  \includegraphics[width=\figwidth]{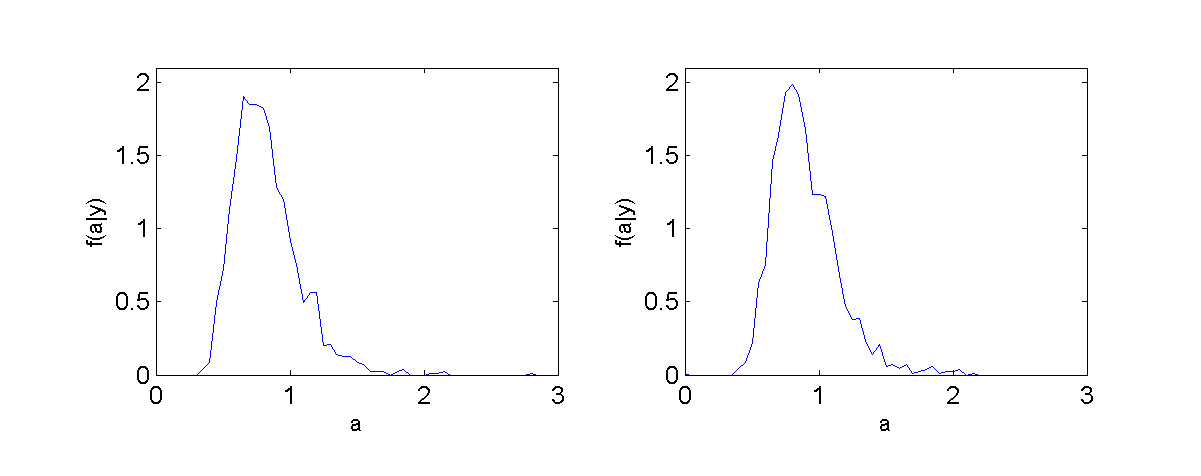}
  \caption{Posterior distribution of hyperparameter $a$ (left: $\mathrm{SNR} = 2$dB, right: $\mathrm{SNR} = 20$dB).}
  \label{fig:hist_a}
\end{figure}

\begin{figure}[h!]
  \centering
  \includegraphics[width=\figwidth]{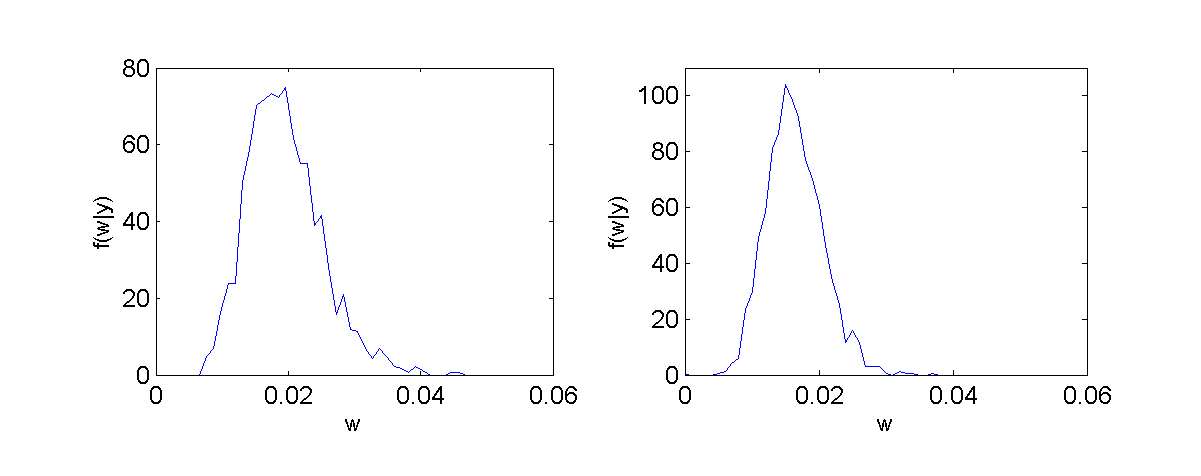}
  \caption{Posterior distribution of hyperparameter $w$ (left: $\mathrm{SNR} = 2$dB, right: $\mathrm{SNR} = 20$dB).}
  \label{fig:hist_w}
\end{figure}

\begin{figure}[h!]
  \centering
  \includegraphics[width=\figwidth]{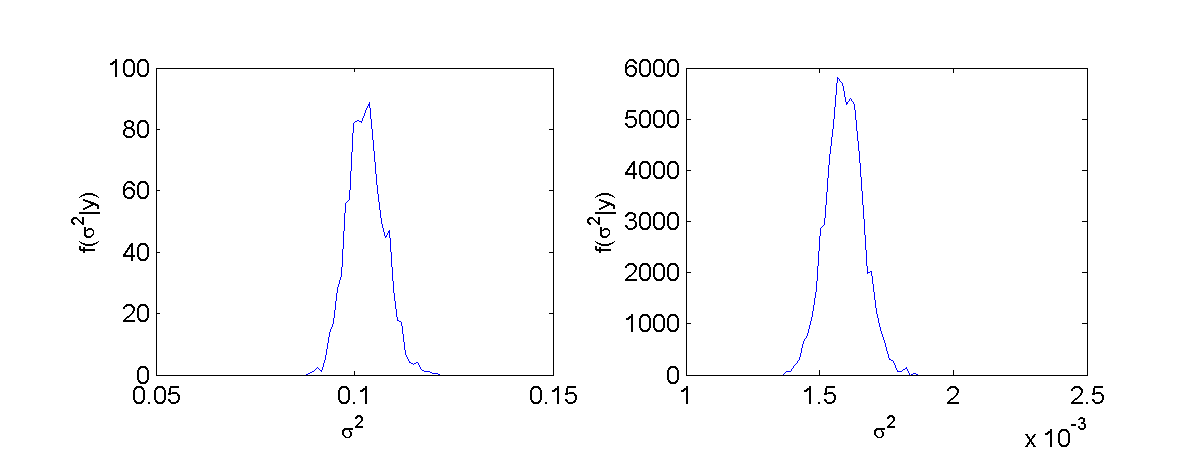}
  \caption{Posterior distribution of hyperparameter $\noisevar$ (left: $\mathrm{SNR} = 2$dB, right: $\mathrm{SNR} = 20$dB).}
  \label{fig:hist_sigma2}
\end{figure}

Moreover, as the proposed algorithm generates samples distributed
according to the posterior distribution in \eqref{eq:posterior},
these samples can be used to compute the posterior distributions of
each parameter. For illustration, the posterior distributions of the
hyperparameters $a$ and $w$, as well as the noise variance
$\sigma^2$, are shown in Fig.~\ref{fig:hist_a}, \ref{fig:hist_w} and
\ref{fig:hist_sigma2}. These estimated distributions are in good
agreement with the ground truth values of these parameters, randomly
drawn from the prior distribution.

\begin{figure}[h!]
  \centering
  \includegraphics[width=\figwidth]{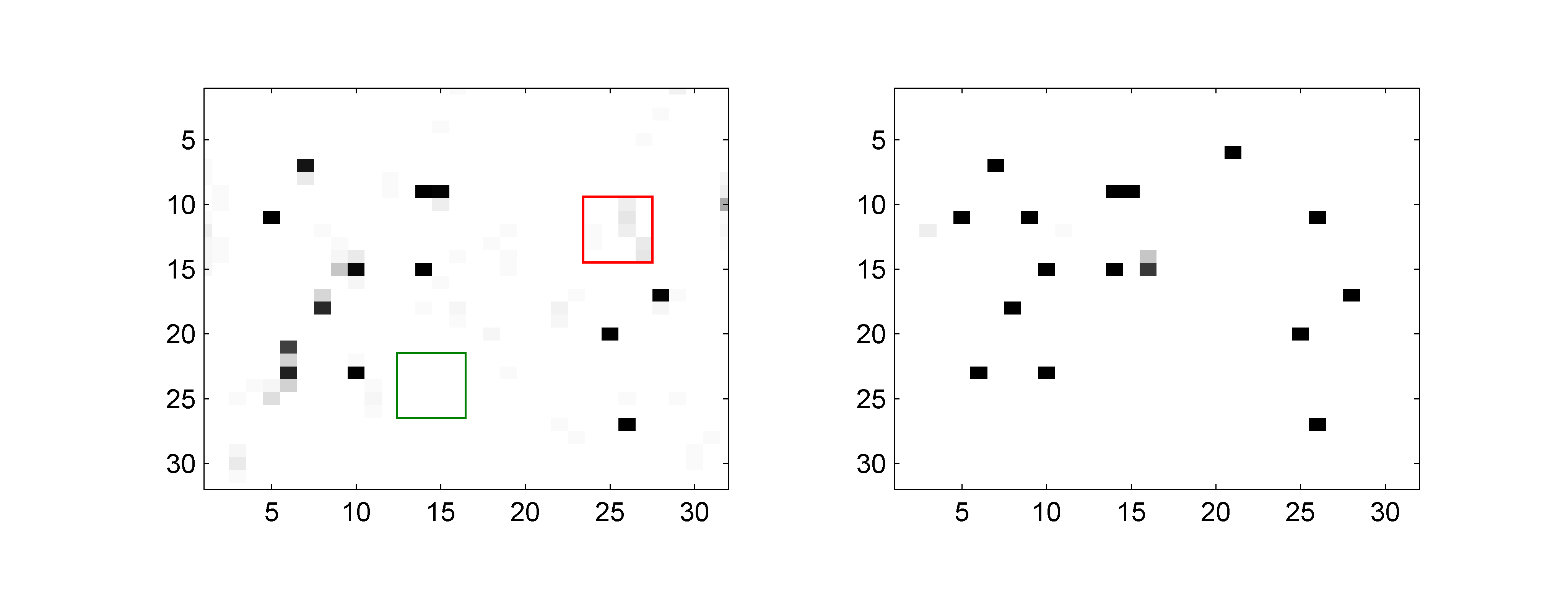}
  \caption{Posterior probabilities of having non-zero pixels (left: $\mathrm{SNR} = 2$dB, right: $\mathrm{SNR} = 20$dB). The probability of having at least
    one non-zero pixel in the red (resp. green) box-delimited area is $45\%$ (resp. $5\%$).
      }\label{fig:indicator_probas}
\end{figure}

In many applications, a measure of confidence that a given pixel or
pixel region is non-zero is of interest. Our Bayesian approach can
easily generate such measures of confidence in the form of posterior
probabilities of the specified event, sometimes known as the
Bayesian p-value. Following the strategy detailed in Appendix
\ref{app:Gene_BeTrG}, the proposed Gibbs sampler generates a
collection of samples
$\left\{\Vima^{(t)}\right\}_{t=1,\ldots,N_{\mathrm{MC}}}$,
distributed according the posterior Bernoulli-truncated Gaussian
distribution in \eqref{eq:posterior_ima}. This sampling requires the
generation of indicator variables $z_i$ ($i=1,\ldots,\dimima$) that
reflect the presence or the absence of non-zero pixel values. It is
the indicator variable $z_i$ that $x_i>0$ that provides information
about non-zero pixels in the image. Using the equivalences
$\left\{z_i=0\right\}\Leftrightarrow \left\{x_i=0\right\}$ and
$\left\{z_i=1\right\}\Leftrightarrow \left\{x_i>0\right\}$, the
posterior probability $\mathrm{P}\left[\ima{i}>0|\Vobs,
\Valpha\right]$ can be easily obtained by averaging over the Gibbs
samples of the binary variables
$\left\{z_i^{(t)}\right\}_{t=N_{\textrm{bi}+1},\ldots,N_{\textrm{MC}}}$.
To illustrate, these probabilities are depicted in
Fig.~\ref{fig:indicator_probas}. In addition, these Gibbs samples
can be used to compute the probability of having non-zero pixels in
a given area of the image. The estimated posterior probability for
the event that a non-zero pixel is present inside the small red
rectangle in the figure is equal to $45\%$ for the case of
$\mathrm{SNR} = 2$dB. Conversely, the posterior probability of
having a non-zero pixel in the green box is $5\%$. For $\mathrm{SNR}
= 20$dB the MAP algorithm correctly detects up the presence of a
pixel in this region. On the other hand, even though at
$\mathrm{SNR} = 2$dB the MAP reconstruction has not detected this
pixel, we can be $45\%$ confident of the presence of such a pixel in
the red rectangular region on the left panel of Fig.
\eqref{fig:indicator_probas}.

\begin{figure}[h!]
  \centering
  \includegraphics[width=\figwidth]{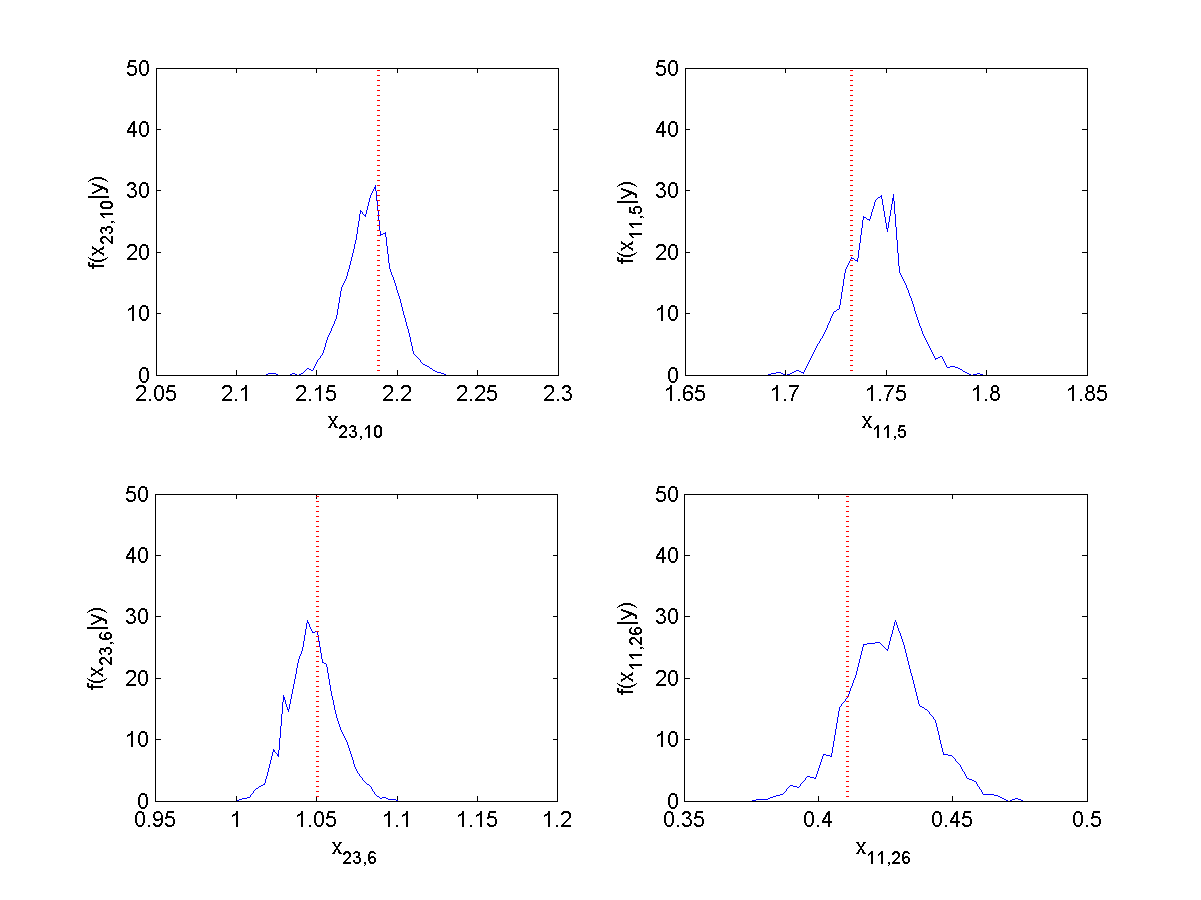}
  \caption{Posterior distributions of the non-zero values of $\Vima$ for $4$ different pixel locations and for $\mathrm{SNR} = 20$dB (actual pixel intensity values are depicted with dotted red lines).}
  \label{fig:hist_theta_nonzero}
\end{figure}

The posterior distributions of four different pixels are depicted in
Fig.~\ref{fig:hist_theta_nonzero}. These posteriors are consistent
with the actual values of these pixels that are represented as
dotted red lines in these figures.  In particular, in all cases the
actual values all lie within the $75\%$ central quantile of the
posterior distribution.

\subsubsection{Comparison of reconstruction performances}
\label{subsec:simu_perf} Here we compare our proposed hierarchical
Bayesian method to the sparse reconstruction methods of
\cite{Ting2006,Ting2006icip}. The techniques proposed in
\cite{Ting2006,Ting2006icip} are based on penalized likelihood EM
algorithms that perform empirical estimation of the unknown
hyperparameters. Therein, two empirical Bayesian estimators, denoted
Emp-MAP-Lap and Emp-MAP-LAZE, based on a Laplacian or a LAZE prior
respectively, were proposed. We also compare to the standard
Landweber algorithm \cite{Landweber1951}  that has been previously
used to perform MRFM image reconstruction
\cite{Zuger1993,Degen2009}. These are compared to our hierarchical
Bayesian MAP reconstruction algorithm, given in
\eqref{eq:MAP_estimate}, and also to a minimum mean square error
(MMSE) reconstruction algorithm extracted from the estimated full
Bayes posterior \eqref{eq:posterior}. The MMSE estimator of the
image $\Vima$ is obtained by empirical averaging over the last
$N_r=1700$ samples of the Gibbs sampler according to:
\begin{equation}
\label{eq:MMSE_estimate}
\begin{split}
    \hat\Vima_{\text{MMSE}} &= \mathrm{E}\left[\Vima|\Vobs\right] \\
                            & \approx \frac{1}{N_r} \sum_{t=1}^{N_r}
\sample{\Vima}{N_\textrm{bi}+t}.
\end{split}
\end{equation}

As in \cite{Ting2006} we compare the various reconstruction
algorithms with respect to several performance criteria. Let
$\bfe=\Vima-\hat\Vima$ denote the reconstruction error when
$\hat\Vima$ is the estimator of the image $\Vima$ to be recovered.
These criteria are: the $\ell_0$, $\ell_1$ and $\ell_2$-norms of
$\bfe$, which measures the accuracy of the reconstruction, and the
$\ell_0$-norm of the estimator $\hat\Vima$, which measures its
sparsity. As pointed out in \cite{Ting2006}, a human observer can
usually not visually detect the presence of non-zero intensities if
they are below a small threshold. Thus, a less strict
measure\footnote{The introduced measure of sparsity is denoted
$\norm{\cdot}_\delta$. This is an abuse of notation since it is not
a norm.} of sparsity than the $\ell_0$-norm, which is denoted
$\norm{\cdot}_\delta$, is the number of reconstructed image pixels
that are less than a given threshold $\delta$:
\begin{equation}
\begin{split}
    \norm{\hat\Vima}_\delta &= \sum_{i=1}^{\dimm}
        \Indicfun{\hat{x}_i<\delta}{\hat{x}_i},\\
    \norm{\bfe}_\delta &= \sum_{i=1}^{\dimm}
        \Indicfun{e_i<\delta}{e_i}.
\end{split}
\end{equation}
It what follows, $\delta$ has been chosen as $\delta=10^{-2}
\norm{\Vima}_\infty$. To summarize, the following criteria have been
computed for the image in paragraph~\ref{subsec:simu_results} for
two levels of SNR: $\norm{e}_0$, $\norm{e}_\delta$, $\norm{e}_1$,
$\norm{e}_2$, $\norm{\hat\Vima}_0$ and $\norm{\hat\Vima}_\delta$.

Table~\ref{tab:performance} shows the six performance measures for
the five different algorithms studied. The proposed Bayesian methods
(labeled ``proposed MMSE" and ``proposed MAP" in the table)
outperform the other reconstruction algorithms in terms of $\ell_1$
or $\ell_2$-norms. Note that the MMSE estimation of the unknown
image is a non sparse estimator in the $\ell_0$-norm sense. This is
due to the very small but non-zero posterior probability of non-zero
value at many pixels. The sparsity measure $\norm{\cdot}_\delta$
indicates that most of the pixels are in fact very close to zero.
The MAP reconstruction method seems to achieve the best balance
between the sparsity of the solution and the minimization of the
reconstruction error. Of course, by its very construction, the MMSE
reconstruction will always have lower mean square error.

\begin{table}[h!]
\renewcommand{\arraystretch}{1.4}
\begin{center}
\caption{Reconstruction performances for different sparse image
reconstruction algorithms.\label{tab:performance}}
\begin{tabular}{|l|r r r r r r|}
\hline
\multirow{2}{5mm}{Method}        & \multicolumn{6}{|c|}{Error criterion}\\
\cline{2-7}
  & $\norm{e}_0$ & $\norm{e}_\delta$ & $\norm{e}_1$ & $\norm{e}_2$  & $\norm{\hat\Vima}_0$  & $\norm{\hat\Vima}_\delta$\\
\hline
\multicolumn{7}{|c|}{$\textrm{SNR}=2$dB}\\
\hline
Landweber          &     1024        &    990     &   339.76    &    13.32      &   1024     &    990\\
Emp-MAP-Lap        &       18        &     17     &    14.13    &     4.40      &      \textbf{0}     &      \textbf{0}\\
Emp-MAP-LAZE       &       60        &     58     &     9.49    &     1.44      &     55     &     55\\
Proposed MMSE      &     1001        &     30     &     3.84    &     \textbf{0.72}      &   1001     &     27\\
Proposed MAP        &       \textbf{19 }       &     \textbf{16}     &    \textbf{2.38}    &     0.81      &     13     &     13\\
\hline
\multicolumn{7}{|c|}{$\textrm{SNR}=20$dB}\\
\hline
Landweber         &      1024      &    931     &    168.85        &     6.67         & 1024  & 931\\
Emp-MAP-Lap       &        33      &     18     &      1.27        &     0.31         &   28  &  23\\
Emp-MAP-LAZE      &       144      &     19     &      1.68        &     0.22         &  144  &  27\\
Proposed MMSE      &       541      &      \textbf{5}     &      \textbf{0.36}        &     \textbf{0.11}         &  541  &  \textbf{16}\\
Proposed MAP       &        \textbf{19}      &     7     &      0.39        &     0.13         &   \textbf{16}  &  \textbf{16}\\
\hline
\end{tabular}
\end{center}
\end{table}

\subsection{Reconstruction of undersampled 3-dimensional images}
\begin{figure*}
  \centering
  \includegraphics[width=\textwidth]{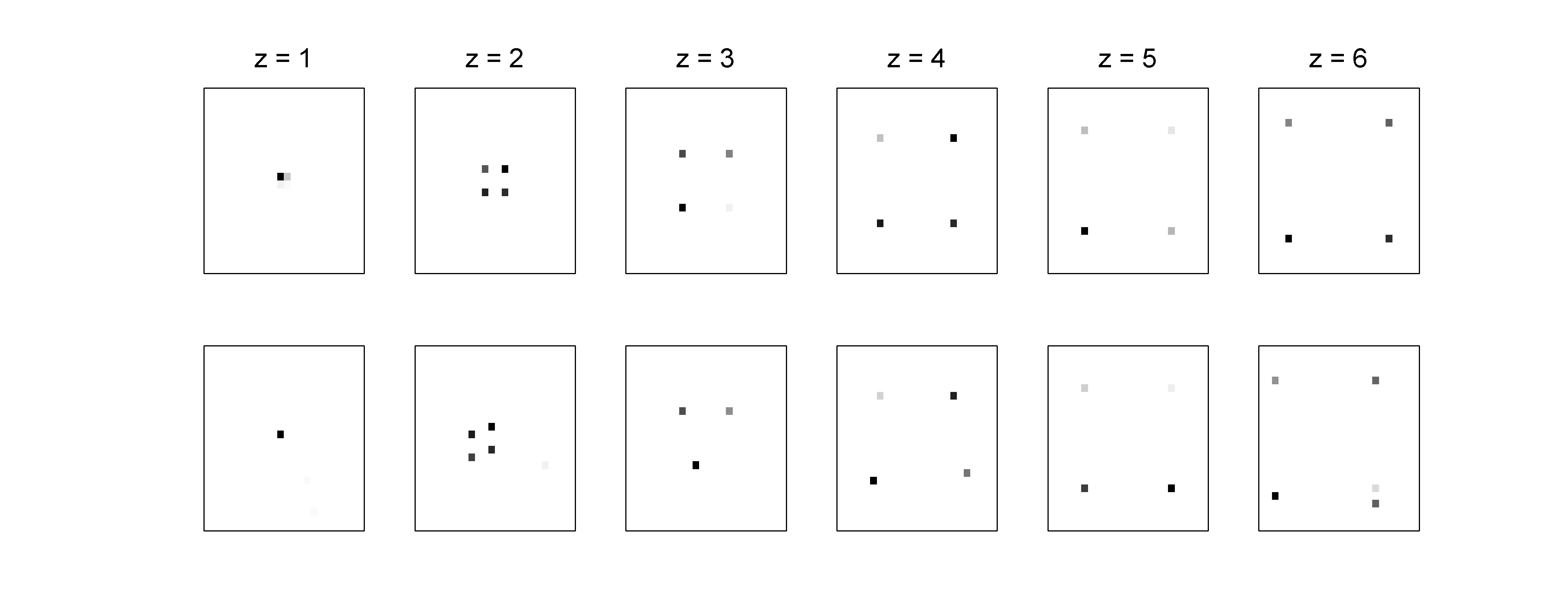}
  \caption{Top: slices of the sparse image to be recovered. Bottom: slices of the estimated sparse image.}
  \label{fig:image_3D_results}
\end{figure*}

\begin{figure}[h!]
  \centering
  \includegraphics[width=\figwidth]{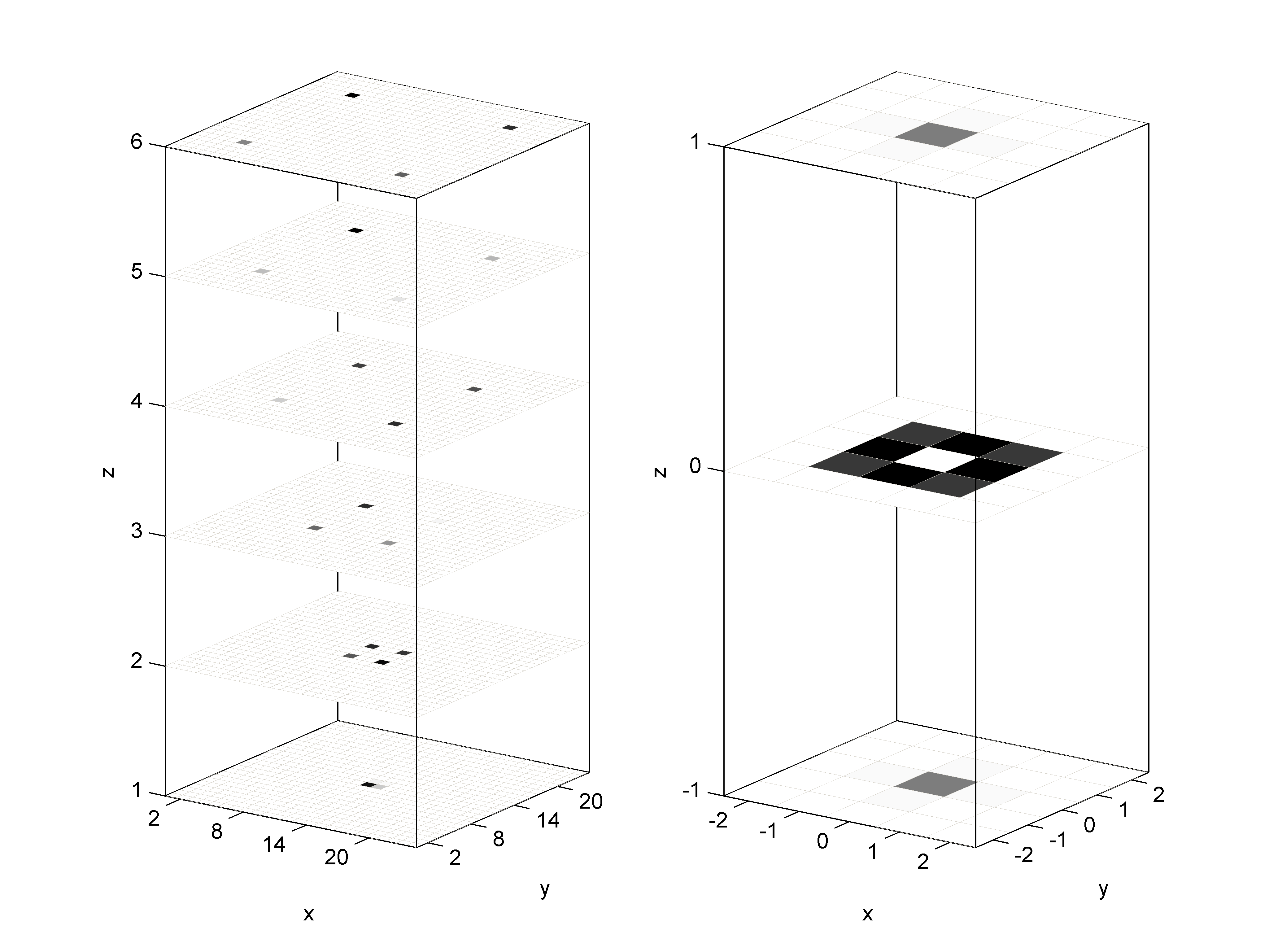}
  \caption{Left: $24\times24\times6$ unknown image to be recovered. Right: $5\times5\times3$ kernel modeling the psf.}
  \label{fig:image_synth_3D}
\end{figure}

\begin{figure}[h!]
  \centering
  \includegraphics[width=\figwidth]{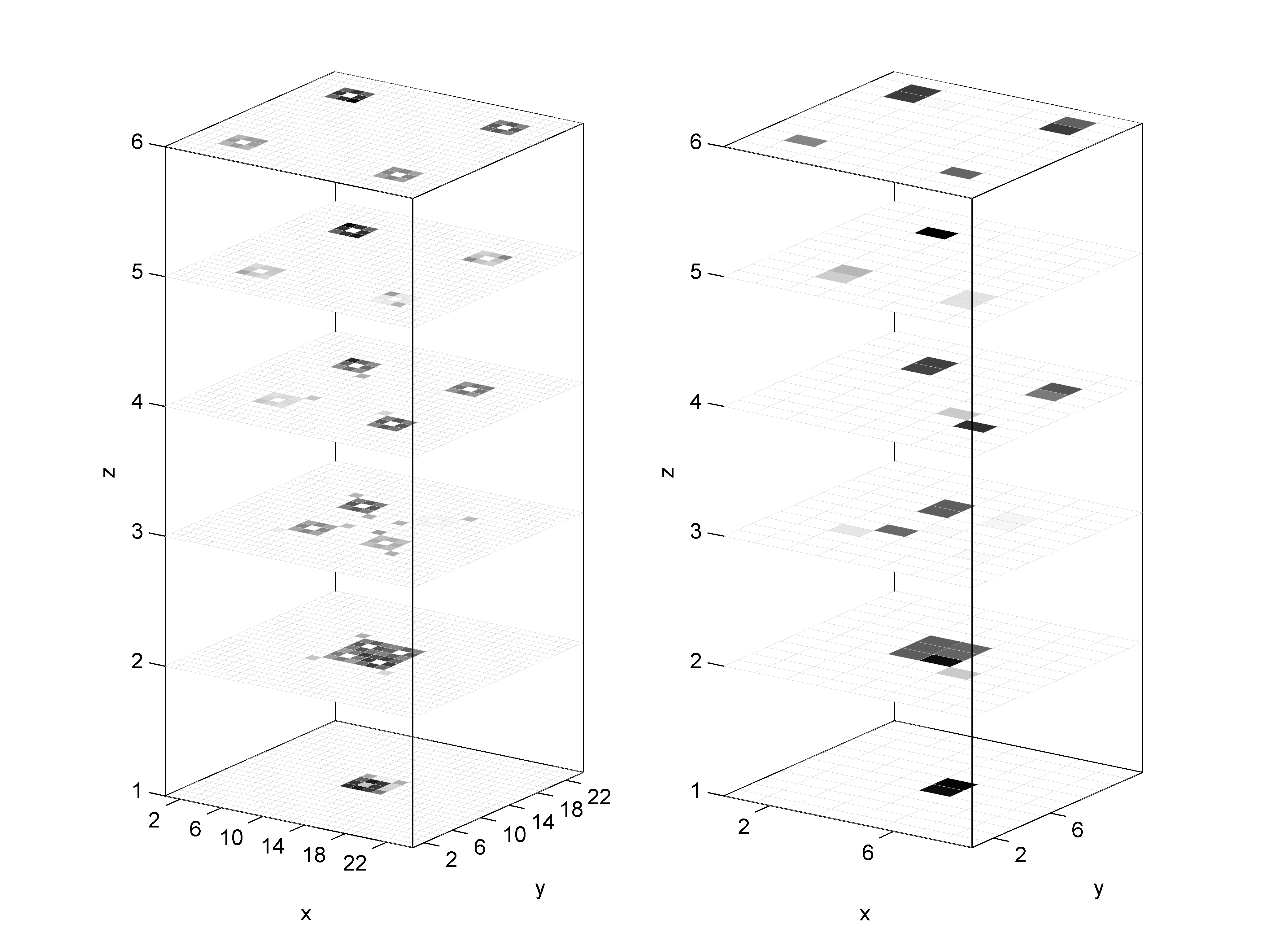}
  \caption{Left: $24\times24\times6$ regularly sampled convolved image. Left: $12\times8\times6$ undersampled observed image.}
  \label{fig:image_observations}
\end{figure}

As discussed below in Section~\ref{sec:simu_real}, the prototype IBM
MRFM instrument \cite{Degen2009} collects data projections as
irregularly spaced, or undersampled, spatial samples. In this
subsection, we indicate how the image reconstruction algorithm can
be adapted to this undersampled scenario in 3D. We illustrate by a
concrete example. First, a $24\times24\times6$ image is generated
such that $4$ pixels have non-zero values in each $z$ slice. The
resulting data is depicted in Fig. \ref{fig:image_3D_results} (top)
and Fig. \ref{fig:image_synth_3D} (left). This image to be recovered
is assumed to be convolved with a $5\times5\times3$ kernel that is
represented in Fig. \ref{fig:image_synth_3D} (right). The resulting
convolved image is depicted in Fig. \ref{fig:image_observations}
(left). However, the actual observed image is an undersampled
version of this image. More precisely, the sampling rates are
assumed to be $d_x=2$, $d_y=3$, $d_z=1$, respectively, in the $3$
dimensions. Consequently the observed $3$D image, shown in Fig.
\ref{fig:image_observations}, is of size $12\times8\times6$.
Finally, an i.i.d. Gaussian noise with $\sigma=0.02$ is added
following the model in \eqref{eq:model_nD}. Note that under these
assumptions, the application $\ftrans{\cdot}{\cdot}$ can be split
into two standard operations following the composition:
\begin{equation}
  \ftrans{\psf}{\bfX} = g_{d_x,d_y,d_z}\left(\psf \otimes
\bfX\right),
\end{equation}
where $g_{d_x,d_y,d_z}\left(\cdot\right)$ stands for the
undersampling function.

The proposed hierarchical Bayesian algorithm is used to perform the
sparse reconstruction with undersampled data. The number of Monte
Carlo runs was fixed to $N_\textrm{MC}=2000$ with
$N_\textrm{bi}=300$ burn-in iterations. Figure
\ref{fig:image_3D_results} shows the result of applying the proposed
MAP estimator to the estimated posterior.

\section{Application on real MRFM images}
\label{sec:simu_real} 
\begin{figure*}
  \centering
  \includegraphics[width=\textwidth]{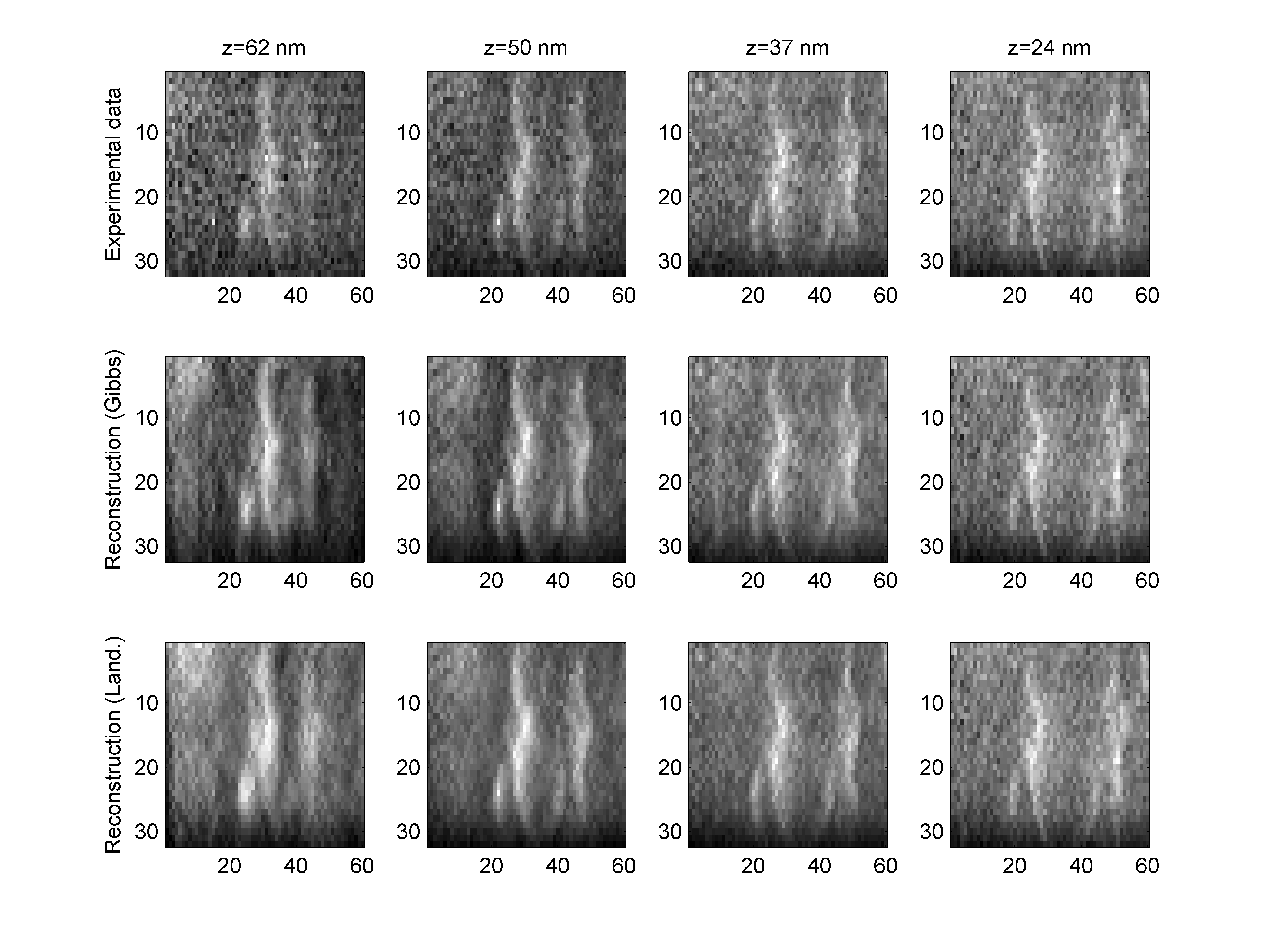}
  \caption{Top: experimental scan data where black (resp. white) pixel represents low (resp. high) density of spin (as in \cite{Degen2009}). Middle: scan data reconstructed from the proposed hierarchical Bayesian algorithm.
  Bottom: scan data reconstructed from the Landweber algorithm.}
  \label{fig:real_data_4slices}
\end{figure*}

Here we illustrate the hierarchical Bayesian
MAP reconstruction algorithm for real three dimensional MRFM data.
The data is a set of MRFM projections of a sample of tobacco virus.
Comprehensive details of both the experiment and the MRFM data
acquisition protocol are given in \cite{Degen2009} and the
supplementary materials \cite{Degen2009compl}. The observed sample
consists of a collection of Tobacco mosaic virus particles that
constitute a whole viral segment in addition to viral fragments. The
projections are computed from the measured proton distribution and
the $3$-dimensional psf following the protocol described in
\cite{Degen2009} and \cite{Degen2009compl}. The resulting scan data
are depicted in Figure~\ref{fig:real_data_4slices} (top) for four
different distances between the MRFM tip and the sample: $d=24$nm,
$d=37$nm, $d=50$nm and $d=62$nm. Each of these x-y slices is of size
$60 \times 32$ pixels.

\begin{figure}[h!]
  \centering
  \includegraphics[width=\figwidth]{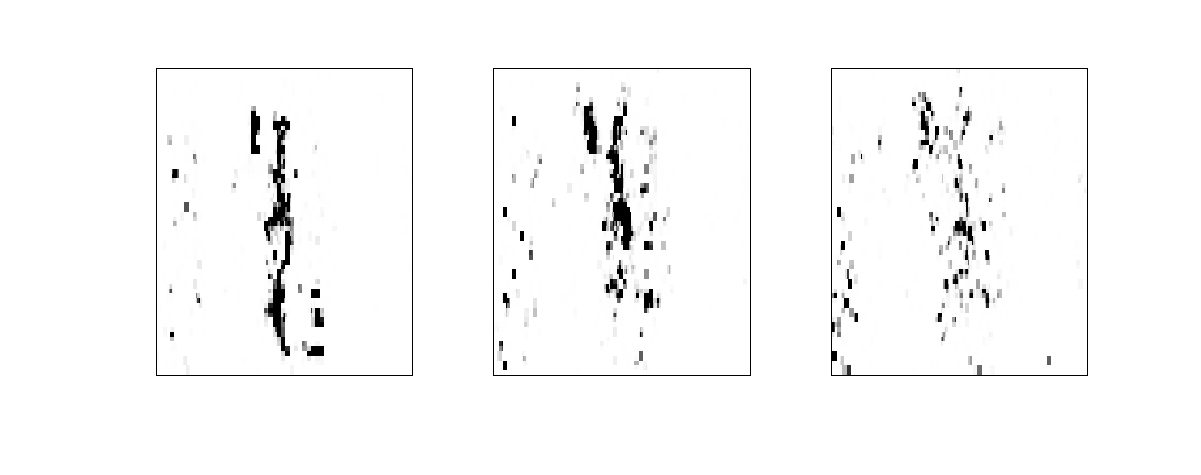}
  \caption{Three horizontal slices of the estimated image.}
  \label{fig:real_3_estimated_slices}
\end{figure}

\begin{figure}[h!]
  \centering
  \includegraphics[width=\figwidth]{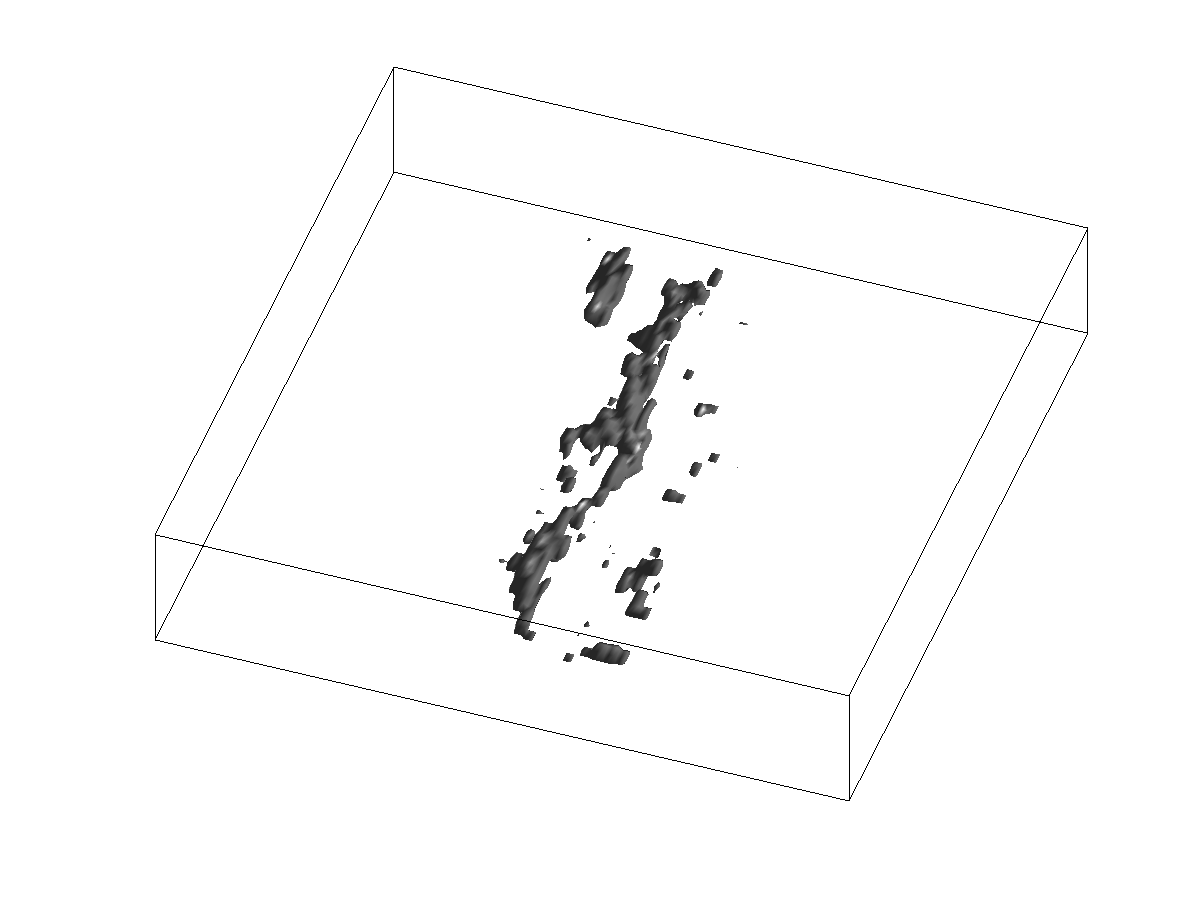}
  \caption{$3$-dimensional view of the estimated profile of the Tobacco virus fragments.}
  \label{fig:result_data_Rugar_MAP_3D}
\end{figure}

These experimental data are undersampled, i.e. the spatial
resolution of the MRFM tip, and therefore the psf function, is finer
than the resolution of the observed slices. Consequently, these data
have been deconvolved taking into account the oversampling rates
defined by $d_x=3$, $d_y=2$ and $d_z=3$ in the three directions. The
MAP estimate of the unknown image is computed from
$N_{\textrm{MC}}=1000$ Gibbs samples of the proposed Bayesian
algorithm initialized with the output of a single Landweber
iteration. Several more iterations of the Landweber algorithm would
produce the reconstructions reported in \cite{Degen2009}. Three
horizontal slices of the estimated image\footnote{Note that most
part of the estimated $3$ dimensional image is empty space due to
the very localized proton spin centers in the image.} are depicted
in Figure~\ref{fig:real_3_estimated_slices}. A $3$-dimensional view
of the estimated profile of the virus fragments is shown in
Figure~\ref{fig:result_data_Rugar_MAP_3D}. The MMSE estimates of the
parameters introduced in Section~\ref{sec:model} are
$\hat{\sigma}^2_{\textrm{MMSE}} = 0.10$, $\hat{a}_{\textrm{MMSE}} =
1.9\times 10^{-12}$ and $\hat{w}_{\textrm{MMSE}} = 1.4 \times
10^{-2}$. The image reconstructions produced by the Landweber and
Bayesian MAP algorithms are shown in Fig.
\ref{fig:real_data_4slices}.

\begin{figure}[h!]
  \centering
  \includegraphics[width=\figwidth]{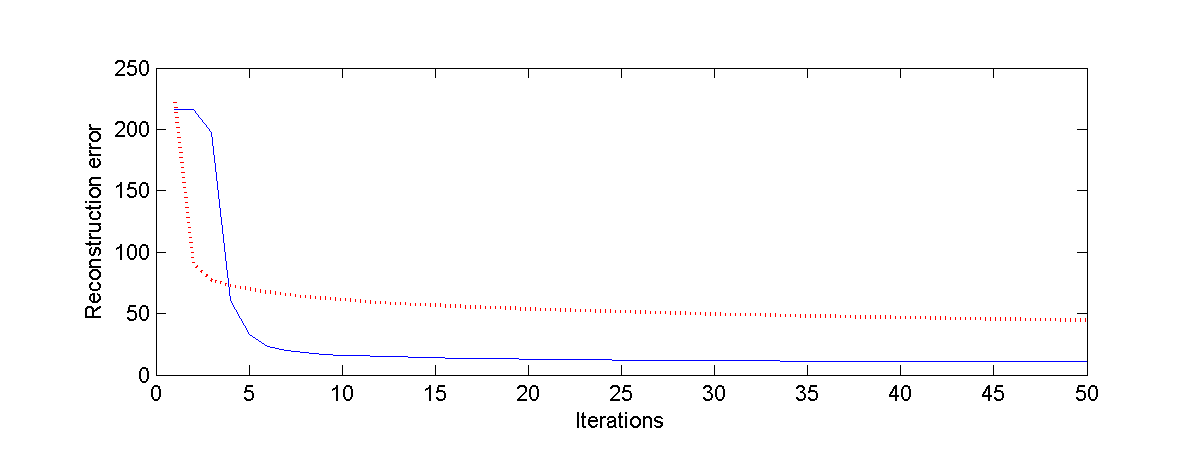}
  \caption{Error of the reconstructions as functions of the iteration number for the
    proposed algorithm (continuous blue line) and Landweber algorithm (dotted red line).}
  \label{fig:real_reconstruction_error}
\end{figure}

By forward projecting the estimated virus image through the point
spread function one can visually evaluate the goodness of fit of the
reconstruction to the raw measured data. This is depicted in
Fig.~\ref{fig:real_data_4slices}. These figures are clearly in good
agreement with the observed data (top). To evaluate the convergence
speed, the reconstruction error is represented in
Figure~\ref{fig:real_reconstruction_error} as a function of the
iterations for the proposed Bayesian and the Landweber algorithms.
This shows that the convergence rate of our algorithm is
significantly better than the Landweber algorithm.

\section{Conclusions}
\label{sec:conclusions} This paper presented a hierarchical Bayesian
algorithm for deconvolving sparse positive images corrupted by
additive Gaussian noise. A Bernoulli-truncated exponential
distribution was proposed as a prior for the sparse image to be
recovered. The unknown hyperparameters of the model were integrated
out of the posterior distribution of the image, producing a full
posterior distribution that can be used for estimation of the pixel
values by extracting the mode (MAP) or the first moment (MMSE). An
efficient Gibbs sampler was used to generate approximations to these
estimates. The derived Bayesian estimators significantly
outperformed several previously proposed sparse reconstruction
algorithms. Our approach was implemented on real MRFM data to
reconstruct a $3$D image of a tobacco virus. Future work will
include extension of the proposed method to other sparse bases,
inclusion of uncertain point spread functions, and investigation of
molecular priors. Future investigations might also include a
comparison between the proposed MCMC approach and variational Bayes
approaches.

 \appendices

\section{Derivation of the conditional \\posterior distribution $f\left(\ima{i} \left| w,a,\noisevar,\Vima_{-i},\Vobs\right.\right)$}
\label{app:derivation1}

The posterior distribution of each component $\ima{i}$
($i=1,\ldots,\dimm$) conditionally upon the others is linked to the
likelihood function \eqref{eq:likelihood} and the prior distribution
\eqref{eq:prior_ima2} via the Bayes' formula:
\begin{equation}
    \label{eqApp:posterior_ima1}
f\left(\ima{i} | w,a,\noisevar,\Vima_{-i},\Vobs\right) \propto
f\left(\Vobs | \Vima,\noisevar \right) f\left(\ima{i} | w,a\right).
\end{equation}
This distribution  can be easily derived by decomposing $\Vima$ on
the standard orthonormal basis
 \begin{equation}
   \mathbb{B}
=\left\{{\bf{u}}_1,\ldots,{\bf{u}}_\dimm\right\},
\end{equation}
where ${\bf{u}}_i$ is the $i$th column of the $\dimm \times \dimm$
identity matrix. Indeed, let decompose $\Vima$ as follows:
\begin{equation}
 \label{eqApp:decomposition}
  \Vima = \tilde{\Vima}_i + x_i{\bf{u}}_i,
\end{equation}
where $\tilde{\Vima}_i$ is the vector $\Vima$ whose $i$th element
has been replaced by $0$. Then the linear property of the operator
$\ftrans{\psf}{\cdot}$ allows one to state:
\begin{equation}
\ftrans{\psf}{\Vima} = \ftrans{\psf}{\tilde{\Vima}_i} +
x_i\ftrans{\psf}{{\bf{u}}_i}.
\end{equation}
Consequently, \eqref{eqApp:posterior_ima1} can be rewritten
\begin{equation}
\begin{split}
  f&\left(\ima{i} | w,a,\noisevar,\Vima_{-i},\Vobs\right) \propto
\exp\left(-\frac{\norm{\bfe_i -
\ima{i}\Vtrans{i}}^2}{2\noisevar}\right)\\
&\times \left[(1-w)\delta\left(\ima{i}\right) +
\frac{w}{a}\exp\left(-\frac{\ima{i}}{a}\right)\Indicfun{\R_+^*}{\ima{i}}\right],
\end{split}
\end{equation}
where\footnote{It can be noticed that, for deblurring applications,
$\Vtrans{i}$ is also the $i$th column of the matrix $\MATtrans$
introduced in \eqref{eq:model}.}
\begin{equation}
\label{eqApp:useful} \left\{
  \begin{split}
    \bfe_i &= \Vobs-\ftrans{\psf}{\tilde{\Vima}_i},\\
    \Vtrans{i} &=\ftrans{\psf}{{\bf{u}}_i}.
  \end{split}
\right.
\end{equation}
An efficient way to compute $\bfe_i$ within the Gibbs sampler scheme
is reported in Appendix~\ref{app:computations}. Then,
straightforward computations similar to those in \cite{Cheng1996}
and \cite[Annex B]{Mazet2005phd} yield to the following
distribution:
\begin{equation}
\begin{split}
    \label{eqApp:posterior_ima2}
  f\left(\ima{i} | w,a,\noisevar,\Vima_{-i},\Vobs\right) &\propto
(1-w_i) \delta\left(\ima{i}\right) \\
&+ w_i \phi_+\left(\ima{i}|\mu_i,\eta^2_i\right),
\end{split}
\end{equation}
with
\begin{equation}\label{eqApp:mean_var_posterior_ima}
\left\{
  \begin{split}
    &\eta^2_i = \frac{\noisevar}{\norm{\Vtrans{i}}^2}, \\
    &\mu_i = \eta^2_i
            \left(\frac{\Vtrans{i}\transp\bfe_i}{\noisevar}-\frac{1}{a}\right),
  \end{split}\right.
\end{equation}
and
\begin{equation}\label{eqApp:proba_posterior_ima}
\left\{
  \begin{split}
   &u_i = \frac{w}{a} C\left(\mu_i,\eta_i^2\right)
             \exp\left(\frac{\mu_i^2}{2\eta^2_i}\right),\\
   &w_i = \frac{u_i}{u_i + (1-w)}.
  \end{split}\right.
\end{equation}
The distribution in \eqref{eqApp:posterior_ima2} is a
Bernoulli-truncated Gaussian distribution with hidden mean $\mu_i$
and hidden variance $\eta^2_i$.

\section{Fast recursive computations \\ for simulating  according to $f\left(\Vima \left| w,a,\noisevar,\Vobs\right.\right)$}
 \label{app:computations}
In the Gibbs sampling strategy presented in Section~\ref{sec:Gibbs},
the main computationally expensive task is the generation of samples
distributed according to $f\left(\ima{i} \left|
w,a,\noisevar,\Vima_{-i},\Vobs\right.\right)$. Indeed, the
evaluation of the hidden mean and hidden variance in
\eqref{eqApp:mean_var_posterior_ima} of the Bernoulli-truncated
Gaussian distribution may be costly, especially when the bilinear
application $\ftrans{\cdot}{\cdot}$ is not easily computable. In
this appendix, an appropriate recursive strategy is proposed to
accelerate the Gibbs sampling by efficiently updating the coordinate
$i$ of the vector $\Vima$ at iteration $t$ of the Gibbs sampler.

Let $\samplebis{\Vima}{t,i-1}$ denote the current Monte Carlo state
of the unknown vectorized image $\Vima$ ($i=1,\ldots,\dimm$):
\begin{equation}
  \samplebis{\Vima}{t,i-1} =
\left[\ima{1}^{(t)},\ldots,\ima{i-1}^{(t)},\ima{i}^{(t-1)},\ima{i+1}^{(t-1)},\ldots,\ima{\dimm}^{(t-1)}\right]\transp.
\end{equation}
with, by definition,
$\samplebis{\Vima}{t,0}=\samplebis{\Vima}{t-1,M}$. Updating $
\samplebis{\Vima}{t,i-1}$ consists of drawing $\ima{i}^{(t)}$
according to the Bernoulli-truncated Gaussian distribution
$f\left(\ima{i} \left|
w,a,\noisevar,\samplebis{\Vima}{t,i-1}_{-i},\Vobs\right.\right)$ in
\eqref{eq:posterior_ima} with:
\begin{equation}
  \samplebis{\Vima}{t,i-1}_{-i} =
\left[\ima{1}^{(t)},\ldots,\ima{i-1}^{(t)},\ima{i+1}^{(t-1)},\ldots,\ima{\dimm}^{(t-1)}\right]\transp.
\end{equation}
The proposed strategy to simulate efficiently according to
\eqref{eq:posterior_ima} is based on the following property.\\

\emph{Property}: Given the quantity
$\ftrans{\psf}{\sample{\Vima}{0}}$ and the vectors
$\left\{\bfh_i\right\}_{i=1,\ldots,\dimm}$, simulating according to
$f\left(\ima{i} \left|
w,a,\noisevar,\samplebis{\Vima}{t,i}_{-i},\Vobs\right.\right)$ can
be performed without evaluating the bilinear
function $\ftrans{\cdot}{\cdot}$.\\

\emph{Proof}: Simulating according to \eqref{eq:posterior_ima}
mainly requires to compute the vector $\bfe_i$ introduced by
\eqref{eqApp:useful}:
\begin{equation}
    \label{eqApp:error}
  \bfe_i = \Vobs - \ftrans{\psf}{\tilde{\Vima}^{(t,i-1)}_i},
\end{equation}
with
\begin{equation}
  \tilde{\Vima}^{(t,i-1)}_i =
\left[\ima{1}^{(t)},\ldots,\ima{i-1}^{(t)},0,\ima{i+1}^{(t-1)},\ldots,\ima{\dimm}^{(t-1)}\right]\transp.
\end{equation}
Moreover, by using the decomposition in \eqref{eqApp:decomposition}
and by exploiting the linear property of $\ftrans{\psf}{\cdot}$, the
vector $\ftrans{\psf}{\tilde{\Vima}^{(t,i-1)}_i}$ in the right-hand
side of \eqref{eqApp:error} can be rewritten as:
\begin{equation}
\label{eqApp:recursion_bis}
  \ftrans{\psf}{\tilde{\Vima}^{(t,i-1)}_i} = \ftrans{\psf}{{\Vima}^{(t,i-1)}} -
\ima{i}^{(t-1)}\bfh_i,
\end{equation}
where $\bfh_i$ has been introduced in \eqref{eqApp:useful}.
Consequently, to prove the property, we have to demonstrate that the
vector series
$\left\{\ftrans{\psf}{{\Vima}^{(t,k)}}\right\}_{k=1,\ldots,\dimm}$
can be computed recursively without using $\ftrans{\cdot}{\cdot}$.
Assume that $\ftrans{\psf}{{\Vima}^{(t,i-1)}}$ is available at this
stage of the Gibbs sampling and that $\ima{i}^{(t)}$ has been drawn.
The new Monte Carlo state is then:
\begin{equation}
  \samplebis{\Vima}{t,i} =
\left[\ima{1}^{(t)},\ldots,\ima{i-1}^{(t)},\ima{i}^{(t)},\ima{i+1}^{(t-1)},\ldots,\ima{\dimm}^{(t-1)}\right]\transp.
\end{equation}
Similarly to \eqref{eqApp:recursion_bis}, the vector
$\ftrans{\psf}{\Vima^{(t,i)}}$ can be decomposed as follows:
\begin{equation}
\label{eqApp:recursion}
  \ftrans{\psf}{\Vima^{(t,i)}} =
 \ftrans{\psf}{\tilde\Vima^{(t,i-1)}_{i}} + \ima{i}^{(t)} \bfh_i.
\end{equation}
 Therefore, combining  \eqref{eqApp:recursion_bis} and
\eqref{eqApp:recursion}
 allow one to state:
\begin{equation*}
  \ftrans{\psf}{\Vima^{(t,i)}} =
 \ftrans{\psf}{\Vima^{(t,i-1)}}  + \left(\ima{i}^{(t)}  - \ima{i}^{(t-1)}\right)\bfh_i.
\end{equation*}
\begin{flushright}
  \small{$\blacksquare$}
\end{flushright}
The bilinear function $\ftrans{\cdot}{\cdot}$ only needs to be used
at the very beginning of the Gibbs sampling algorithm to evaluate
$\ftrans{\psf}{\sample{\Vima}{0}}$ and the vectors
$\left\{\bfh_i\right\}_{i=1,\ldots,\dimm}$. The resulting simulation
scheme corresponding to step 3 of Algorithm~\ref{algo:Gibbs} is
shown in Algorithm~\ref{algo:efficient_sim}.

\begin{figure}[h!]
\begin{algogo}{Efficient simulation according to  $f\left(\Vima \left|
w,a,\noisevar,\Vobs\right.\right)$}
    \label{algo:efficient_sim}
    For $i=1,\ldots,\dimm,$ update the $i$th coordinate of the vector $$
  \samplebis{\Vima}{t,i-1} =
\left[\ima{1}^{(t)},\ldots,\ima{i-1}^{(t)},\ima{i}^{(t-1)},\ima{i+1}^{(t-1)},\ldots,\ima{\dimm}^{(t-1)}\right]\transp$$
via the following steps:
    \begin{itemize}
        \item[1.] compute $\norm{\bfh_i}^2$,
        \item[2.] set $  \ftrans{\psf}{\tilde{\Vima}^{(t,i-1)}_i} = \ftrans{\psf}{{\Vima}^{(t,i-1)}} -
                \ima{i}^{(t-1)}\bfh_i$,
        \item[3.] set $\bfe_i = \Vima -\ftrans{\psf}{\tilde{\Vima}^{(t,i-1)}_i}$,
        \item[4.] compute $\mu_i$, $\eta^2_i$ and $w_i$ as defined in
            \eqref{eqApp:mean_var_posterior_ima} and \eqref{eqApp:proba_posterior_ima},
        \item[5.] draw $\ima{i}^{(t)}$ according to \eqref{eq:posterior_ima},
        \item[6.] set $\samplebis{\Vima}{t,i} =
                \left[\ima{1}^{(t)},\ldots,\ima{i-1}^{(t)},\ima{i}^{(t)},\ima{i+1}^{(t-1)},\ldots,\ima{\dimm}^{(t-1)}\right]\transp$,
        \item[7.] set $\ftrans{\psf}{{\Vima}^{(t,i)}} = \ftrans{\psf}{\tilde{\Vima}^{(t,i-1)}_i} + \ima{i}^{(t)}\bfh_i$.
    \end{itemize}
\end{algogo}
\end{figure}

\section{Simulation according to a \\Bernoulli-truncated
Gaussian distribution} \label{app:Gene_BeTrG} This appendix
describes how we generate random variables distributed according to
a Bernoulli-truncated Gaussian distribution with parameters
$\left(w_i,\mu_i,\eta_i^2\right)$ whose pdf is:
\begin{equation*}
\begin{split}
  f&\left(x_i|w_i,\mu_i,\eta_i^2\right) = \left(1-w_i\right)\delta(x_i)\\
    & + \frac{w_i}{C\left(\mu_i,\eta_i^2\right)}\exp\left[-\frac{\left(x_i-\mu_i\right)^2}{2\eta_i^2}\right]\Indicfun{\R^*_+}{x_i}
\end{split}
\end{equation*}
where $C\left(\mu_i,\eta_i^2\right)$ has been defined in
\eqref{eq:constant_C}. Monte Carlo draws from this density can be
obtained by using an auxiliary binary variable $z_i$ following the
strategy shown in Algorithm~\ref{algo:Gene_BeTrG}. This indicator
variable takes the value $0$ (resp. $1$) if the pixel $x_i$ is zero
(resp. non-zero).

\begin{figure}[h!]
\begin{algogo}{Simulation according to\\ a Bernoulli-truncated
Gaussian distribution}
    \label{algo:Gene_BeTrG}
    \begin{itemize}
        \item[1.] generate $z_i$ according to  $z_i \sim \calB
            er\left(w_i\right)$,
        \item[2.] set
            $
            \left\{\begin{array}{ll}
                  x_i=0, & \hbox{if $z_i=0$;} \\
                  x_i\sim\calN^+\left(\mu_i,\eta_i^2\right), & \hbox{if $z_i=1$.}
                \end{array}
              \right.
            $
    \end{itemize}
\end{algogo}
\end{figure}

In Algorithm~\ref{algo:Gene_BeTrG}, $\calB er\left(\cdot\right)$ and
$\calN^+\left(\cdot,\cdot\right)$ denote the Bernoulli and the
positive truncated Gaussian distributions respectively. In step $2$,
samples distributed according to the truncated Gaussian distribution
can be generated by using an appropriate accept-reject procedure
with instrumental distributions \cite{Geweke1991, Robert1995,
Mazet2005}.

\section*{Acknowledgements}
The authors would like to thank Michael Ting (Seagate Technology)
for providing the code to generate point spread functions of MRFM
tip and Hichem Snoussi (University of Technology of Troyes) for
interesting suggestions regarding this work. The authors are also
very grateful to Dr. Daniel Rugar who provided the real data used in
Section~\ref{sec:simu_real} as well as a valuable feedback on this
paper.

\bibliographystyle{ieeetranS}
\bibliography{biblio}

\end{document}